\documentclass[a4paper,11pt,preprintnumbers]{article}
\pdfoutput=1
\usepackage{jheppub}
\usepackage{mathrsfs}
\usepackage{amsfonts}
\usepackage{amsmath,amssymb,bm}
\usepackage{mathrsfs}
\usepackage{array}
\usepackage{verbatim}
\usepackage{epsfig}
\usepackage{graphicx}
\usepackage[colorlinks=true,citecolor=blue,linkcolor=blue,urlcolor=blue]{hyperref}
\usepackage[normalem]{ulem}
\usepackage{xcolor}
\usepackage{ulem}
\usepackage{multirow}

\usepackage{graphicx}
\usepackage{subcaption}
\graphicspath{{./figures/}}
\usepackage{dcolumn}
\usepackage{bm}
\usepackage{slashed}
\usepackage{siunitx}
\usepackage{ulem,xpatch}
\usepackage{hyperref}
\usepackage[mathlines]{lineno}
\usepackage{comment}
\usepackage{soul}
\usepackage{xcolor}
\usepackage{xfrac}
\allowdisplaybreaks[4]
\newcommand{\inspire}[1]{[\href{https://inspirehep.net/literature?q=#1}{\sc inSPIRE}]}

\def\bea#1\eea{\begin{align}#1\end{align}}

\newcommand{\bef}{\begin{figure}[h!tb]\centering}
\newcommand{\beft}{\begin{figure}[t]\centering}
\newcommand{\eef}{\end{figure}}

\def\QEQ{{%
    \setbox0\hbox{=}%
    \rlap{\hbox to \wd0{\hss?\hss}}\box0
}}

\title{Semi-inclusive deep inelastic scattering off a tensor-polarized spin-1 target}

\author[a]{Jing~Zhao,}
\author[b,c]{Alessandro~Bacchetta,}
\author[d,e,f]{Shunzo~Kumano,}
\author[a,e]{Tianbo~Liu}
\author[a]{and~Ya-jin~Zhou}

\affiliation[a]{Key Laboratory of Particle Physics and Particle Irradiation (MOE),\\
Institute of Frontier and Interdisciplinary Science, Shandong University,\\
Qingdao, Shandong 266237, China}
\affiliation[b]{Dipartimento di Fisica ``A. Volta," Universit\`a di Pavia,\\ 
via Bassi 6, I-27100 Pavia, Italy}
\affiliation[c]{INFN - Sezione di Pavia,\\ 
via Bassi 6, I-27100 Pavia, Italy}
\affiliation[d]
{Quark Matter Research Center,
    Institute of Modern Physics, Chinese Academy of Sciences,\\
    Lanzhou, 730000, China}
\affiliation[e]{ Southern Center for Nuclear Science Theory,\\
    Institute of Modern Physics, 
    Chinese Academy of Sciences,\\
    Huizhou, 516000, China}
\affiliation[f]
{KEK Theory Center, Institute of Particle and Nuclear Studies, KEK,\\
Oho 1-1, Tsukuba, 305-0801, Japan}

\emailAdd{zhaojingzj@sdu.edu.cn}
\emailAdd{alessandro.bacchetta@unipv.it}
\emailAdd{kumanos@impcas.ac.cn}
\emailAdd{liutb@sdu.edu.cn}
\emailAdd{zhouyj@sdu.edu.cn}

\abstract{We investigate the semi-inclusive deep inelastic scattering off a tensor-polarized spin-1 target, focusing on the production of an unpolarized hadron. The complete differential cross section is expressed in terms of 23 structure functions, which depend on the spin states of the target and the azimuthal modulations of the final-state hadron.
Within the transverse momentum dependent (TMD) factorization framework, we derive the hadronic tensor using quark-quark correlators and quark-gluon-quark correlators up to twist-3.
At tree-level, 21 nonvanishing structure functions are obtained at the leading and subleading twist, expressed as convolutions of TMD parton distribution functions and TMD fragmentation functions.
The measurement of these nonzero structure functions can be utilized to explore the tensor-polarized structure of spin-1 particles, offering insights into their internal dynamics.
}


\arxivnumber{2508.06134}

\begin{document}
\maketitle
\flushbottom

\section{Introduction}
\label{s.intro}

Quantum ChromoDynamics (QCD) is the fundamental theory describing strong interactions. Quarks and gluons (i.e., partons) are the elementary degrees of freedom of QCD and are confined inside hadrons. 
The nonperturbative nature of QCD at low-energy scales makes it challenging to compute the properties of hadrons from first principles.
Understanding the internal structure of nucleons and hadrons has become a crucial challenge in nuclear and particle physics (see, e.g., refs.~\cite{Aidala:2012mv,Accardi:2012qut,Diehl:2015uka,Bacchetta:2016ccz,Metz:2016swz,Anselmino:2020vlp,Anderle:2021wcy,Accardi:2023chb,ParticleDataGroup:2024cfk}). This structure is manifested in 
high-energy reactions involving hadrons, whose experimental outcomes can be interpreted on the basis of QCD factorization (see, e.g., refs.~\cite{Collins:1989gx,Collins:2011zzd}). 

Deep inelastic scattering (DIS) plays a significant role in studying the structure of the nucleon.
The large momentum of the exchanged virtual photon, $q$ with $Q^2=-q^2$, provides the hard scale necessary to apply the factorization framework.
In this case, the cross section for the inclusive DIS process can be approximately expressed as a convolution of the lepton-parton hard scattering and the collinear parton distribution functions (PDFs), $f_q(x)$, which indicates the probability density of finding a parton of flavor $q$ carrying a fraction $x$ of the nucleon momentum.
Analogously to PDFs, fragmentation functions (FFs) are introduced to describe the probability density that a parton $q$ produces a hadron $h$ carrying a fraction $z$ of the parent parton.
Inclusive DIS is characterized by a single hard scale and is not sensitive to the intrinsic motion of partons.
To gain sensitivity to parton transverse momentum, we have to consider Semi-Inclusive DIS (SIDIS), i.e., DIS with an additional hadron detected in the final state. The transverse momentum $P_{h\perp}$ of the produced hadron provides an additional scale. In the region where $P_{h\perp} \ll Q$, Transverse-Momentum-Dependent (TMD) factorization can be applied~\cite{Collins:1981uw,Collins:1981uk,Collins:2011zzd, Rogers:2015sqa} and the cross section can be described in terms of TMD PDFs, $f(x,\bm{p}_T^2)$, and TMD FFs, $D(z,\bm{k}_T^2)$, where $\bm{p}_T$ and $\bm{k}_T$ indicate the transverse momenta of the partons with respect to the nucleon and the produced hadron, respectively.

Most investigations of the internal structure of hadrons focused on nucleons, which have spin $=1/2$. The spin structure of nucleons has received great interest since the European Muon Collaboration discovered that quark spin only accounts for a small fraction of the nucleon spin~\cite{EuropeanMuon:1987isl,EuropeanMuon:1989yki}.
The remaining part of nucleon's spin nowadays is believed to originate from the spin of gluons and the orbital angular momenta of all partons~\cite{Kuhn:2008sy,Deur:2018roz}, which requires a comprehensive understanding of the three-dimensional structure of the nucleon~\cite{Yang:2024drd,Bacchetta:2024yzl}. The polarized SIDIS experiments, such as HERMES~\cite{HERMES:1999ryv,HERMES:2001hbj,HERMES:2002buj,HERMES:2004mhh,HERMES:2005mov,HERMES:2020ifk}, COMPASS~\cite{COMPASS:2005csq,COMPASS:2006mkl,COMPASS:2022jth,COMPASS:2012dmt,COMPASS:2014bze}, and JLab~\cite{JeffersonLabHallA:2011ayy,JeffersonLabHallA:2014yxb}, extend our study of nucleon structure to three-dimensional case.

Spin-1 hadrons and nuclei provide unique opportunities to explore novel effects that are inaccessible with spin-1/2 targets.
Studies of the structure of spin-1 particles are still limited, although much progress has been made over the years~\cite{Bora:1997pi,Edelmann:1997qe,Hino:1999qi,Kumano:2020ijt,Kumano:2024fpr,Frankfurt:1983qs,Hoodbhoy:1988am,Chen:2016moq,Kumano:2021fem,Kumano:2020gfk,Kumano:2019igu,Kumano:2016ude,Ninomiya:2017ggn,Strikman:2017koc,Kaur:2020emh,Keller:2022abm,Kumano:2021xau,Sargsian:2022rmq,Yang:2023zod,Cosyn:2024drt,Qiao:2024bgg,Puhan:2023hio}. 
Compared to spin-1/2 particles, spin-1 particles possess five additional independent tensor-polarized components, as described by their spin density matrix~\cite{Bacchetta:2000jk}.
A complete set of TMD PDFs for a spin-1 particle at leading twist and higher has been derived from the decomposition of the quark-quark correlation function~\cite{Bacchetta:2000jk,Kumano:2020ijt,Kumano:2024fpr}, and can be used to describe both inclusive and semi-inclusive DIS processes with a spin-1 target.\footnote{Similar studies for gluon-gluon correlation function were published in ref.~\cite{Boer:2016xqr}.}
The cross section for inclusive DIS off a spin-1 target can be expressed in terms of four tensor-polarized structure functions $b_{1-4}$ in addition to the ones for the spin-1/2 nucleons~\cite{Frankfurt:1983qs,Hoodbhoy:1988am}.
Among them, $b_1$ and $b_2$, as leading-twist structure functions, satisfying the Callan-Gross-like relation $x b_1=b_2$ in the Bjorken scaling limit~\cite{Close:1990zw}, where $x$ is the Bjorken variable.
Although a sum rule $\int dx\, b_1(x) =0$~\cite{Close:1990zw} can be derived when considering only the valence-quark contributions in the tensor structure, this does not imply that $b_1(x)=0$ for spin-1 hadrons and nuclei.
A systematic analysis of jet production in SIDIS with a spin-1 target has been performed in ref.~\cite{Chen:2020ugq}.

The deuteron is the simplest stable spin-1 target used in experiments and is a weakly bound system of a proton and a neutron.
The so-called nuclear shadowing effect is considered to contribute significantly to the structure function $b_1$ at small $x$~\cite{Nikolaev:1996jy,Bora:1997pi,Edelmann:1997qe}, and pions also play a crucial role in deuteron structure~\cite{Miller:2013hla}.
The HERMES collaboration first measured $b_1$ via DIS off a tensor-polarized spin-1 deuteron and found that $b_1$ exhibits a steep rise as $x\rightarrow0$~\cite{HERMES:2005pon}.
Subsequently, quark and antiquark distributions were extracted from a fit to the HERMES data~\cite{Kumano:2010vz}.
Phenomenological studies based on these experimental data have been carried out in refs.~\cite{Cosyn:2017fbo,Cosyn:2020kwu,Kumano:2019igu}.
The magnitude of $b_1$ and the $x$-dependence of $b_1$ calculated using the conventional convolution model exhibit a large difference from experimental results~\cite{Cosyn:2017fbo}, which may suggest that a new hadron mechanism exists. A possible contribution from a hidden color state, a non-standard quark-gluon configuration, has been proposed to explain this discrepancy~\cite{Miller:2013hla}.

Understanding tensor-polarized PDFs for spin-1 hadrons and nuclei can provide insights into their internal structure.
Recent studies indicate that the calculation of deuteron PDFs on the lattice is very challenging~\cite{Chen:2024rgi}, which makes it necessary to extract the PDFs from experimental observables.
The data points from the HERMES collaboration lie approximately within two standard deviations of zero, necessitating a measurement with higher precision. 
Experiments have been proposed at JLab to measure tensor-polarized structure functions in the inclusive and semi-inclusive DIS processes~\cite{Jlabproposal, Poudel:2025tac, Poudel:2025nof}, thus making it possible to access tensor-polarized TMD PDFs.
It has been also proposed to study the structure of spin-1 hadron in proton-deuteron Drell-Yan processes~\cite{Hino:1998ww,Hino:1999qi,Kumano:1999bt,Kumano:2016ude,Qiao:2024bgg}, which could be carried out by the SpinQuest experiment at Fermilab~\cite{Keller:2022abm,Keller:2020wan}.

In this paper, we present a systematic study of the SIDIS process off a tensor-polarized spin-1 target. The differential cross section can be generally expressed in terms of 23 tensor-polarized structure functions, corresponding to specific angular distributions and target spin states.
Within the kinematic region where the target and detected hadron are nearly back-to-back, we apply the TMD factorization to calculate the structure functions at first subleading twist accuracy (twist-3).
We obtain 21 nonvanishing structure functions, which can be used to study the tensor-polarized TMD PDFs for spin-1 particles.

The paper is organized as follows.
In section~\ref{s.kinematics}, we introduce the relevant kinematic variables and spin states for the SIDIS process.
In section~\ref{s.crosssection}, we derive the general form of the differential cross section in terms of the structure functions.
In section~\ref{s.partonmodel}, we give a complete parametrization of quark-quark and quark-gluon-quark correlators, and use them to calculate the structure functions up to twist-3.
A summary and outlook are presented in section~\ref{s.summary}.

\section{Kinematics and spin states}
\label{s.kinematics}
We consider the SIDIS process off a spin-1 target,
\begin{align}
    	\ell (l) +d(P)\rightarrow \ell(l^{\prime})+h(P_{h})+X(P_{X}),
\end{align}
where $d$ denotes the spin-1 target, such as a deuteron.
Throughout this paper, we focus on the unpolarized hadron production in SIDIS with a longitudinal polarized lepton beam and a tensor-polarized target.
The four momenta of the involved particles are indicated by the variables in parentheses. 
Within one-photon-exchange approximation, the momentum of the exchanged virtual photon $q=l-l^\prime$ with $Q^2 = -q^2$ gives its virtuality and provides a hard scale of the process.
Here the target and produced hadron masses are denoted as $M$ and $M_h$, respectively.

\begin{figure}[ht]
	\centering
	\includegraphics[width=0.7\textwidth]{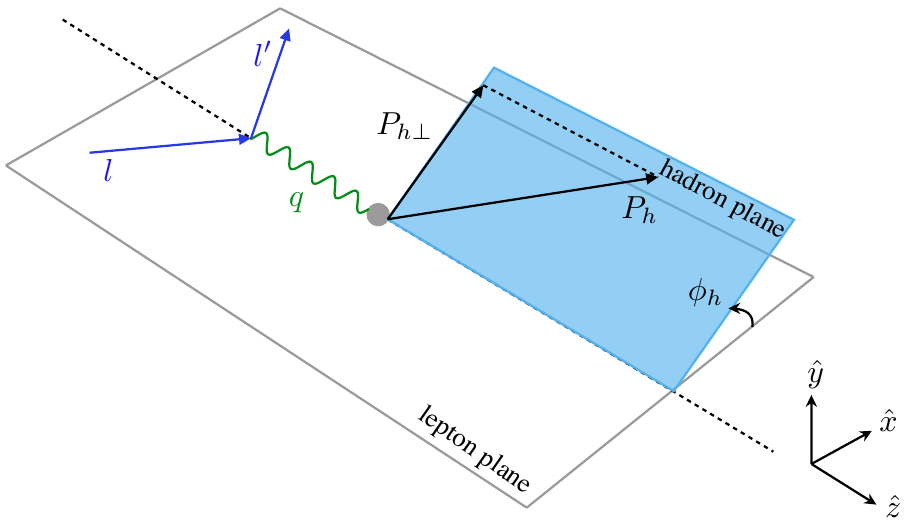}
	\caption{The SIDIS kinematic variables in Trento conventions.}
	\label{f.trento}
\end{figure} 

To describe the SIDIS kinematics, several commonly used dimensionless variables are defined as
\begin{align}
	x_d=\frac{Q^{2}}{2P\cdot q},\quad
    y=\frac{P\cdot q}{P\cdot l},\quad
    z=\frac{P\cdot P_{h}}{P\cdot q},\quad
    \gamma=\frac{2Mx_d}{Q}, \quad \nu=\frac{P\cdot q}{M},
\end{align}
where $x_d$ is the scaling variable for the deuteron. The Bjorken scaling variable, defined as $x_B = Q^2 / (2 M_N \nu)$ with the nucleon mass $M_N$, is related to $x_d$ via $x_B \approx 2x_d$.
Following the Trento conventions~\cite{Bacchetta:2004jz}, we choose the virtual photon-target frame, as illustrated in figure~\ref{f.trento}. 
The azimuthal angle $\phi_h$ spanned by the lepton plane and the hadron plane can be expressed in a Lorentz invariant form as
\begin{align}
	\cos\phi_h=-\frac{l_\mu P_{h\nu} g_\perp^{\mu\nu}}{\sqrt{l_\perp^2 P_{h\perp}^2}},
	\qquad
	\sin\phi_h=-\frac{l_\mu P_{h\nu} \epsilon_\perp^{\mu\nu}}{\sqrt{l_\perp^2 P_{h\perp}^2}},\label{e.phih}
\end{align}
where $l_{\perp}^\mu=g_{\perp}^{\mu \nu} l_\nu$ and $P_{h \perp}^\mu=g_{\perp}^{\mu \nu} P_{h \nu}$ indicate the transverse momenta of the lepton and the detected hadron, respectively. 
We introduce the transverse tensors 
\begin{align}
	g_{\perp}^{\mu\nu} 
	& =g^{\mu\nu}-\frac{q^{\mu}P^{\nu}+q^{\nu}P^{\mu}}{P\cdot q(1+\gamma^{2})}+\frac{\gamma^{2}}{1+\gamma^{2}}\left(\frac{q^{\mu}q^{\nu}}{Q^{2}}-\frac{P^{\mu}P^{\nu}}{M^2}\right),\label{e.gperp}\\
	\epsilon_{\perp}^{\mu\nu} 
	& =\epsilon^{\mu\nu\rho\sigma}\frac{P_{\rho}q_{\sigma}}{P\cdot q\sqrt{1+\gamma^{2}}},\label{e.epsilonperp}
\end{align}
which have nonzero components $g_\perp^{11}=g_\perp^{22}=-1$ and $\epsilon_\perp^{12}=-\epsilon_\perp^{21}=-1$ in the virtual photon-target frame, 
where the convention for totally antisymmetric tensor is $\epsilon^{0123}=1$.

As defined in ref.~\cite{Bacchetta:2000jk}, the polarization for spin-1 particles is characterized by both the spin vector $S^\mu$ and the spin tensor $T^{\mu\nu}$.
We define the transverse and longitudinal spin components with respect to the three-momentum of the hadron.
In light-cone coordinates, the hadron momentum can be written as
\begin{align}
    P^\mu = (P\cdot n) \bar{n}^\mu
    + \frac{M^2}{2P\cdot n} n^\mu,
\end{align} 
where $n$ and $\bar{n}$ are two lightlike basis vectors and satisfy $n\cdot n=0$, $\bar{n}\cdot\bar{n}=0$, and $n\cdot\bar{n}=1$. Then the spin vector and tensor can be parametrized as
\begin{align}
S^{\mu}&= S_{L}\frac{P\cdot n}{M}\bar{n}^{\mu}-S_{L}\frac{M}{2P\cdot n}n^{\mu}+S_{T}^{\mu},\\
T^{\mu\nu}&= \frac12\bigg[\frac43 S_{LL}\frac{(P\cdot n)^2}{M^2}\bar{n}^\mu\bar{n}^\nu-\frac23 S_{LL} (\bar{n}^{\{\mu}n^{\nu\}}-g_T^{\mu\nu})
+\frac13 S_{LL}\frac{M^2}{(P\cdot n)^2}n^\mu n^\nu\nonumber\\
&\quad +\frac{P\cdot n}M\bar{n}^{\{\mu}S_{LT}^{\nu\}} 
-\frac M{2P\cdot n}n^{\{\mu}S_{LT}^{\nu\}}+S_{TT}^{\mu\nu}\bigg],\label{e.spintensor}
\end{align}
where the tensor $g_T^{\mu\nu}$ is given by
\begin{align}
    g_T^{\mu\nu}=g^{\mu\nu}-\bar{n}^\mu n^\nu-n^\mu \bar{n}^\nu.
\end{align}
When $n$ and $\bar{n}$ are chosen as in eq.~\eqref{e.nnbar}, one can obtain the expression of $g_\perp^{\mu\nu}$ in eq.~\eqref{e.gperp}.
The term $a^{\{\mu}b^{\nu\}}$ indicates the symmetrization of the indices, i.e., $a^{\{\mu}b^{\nu\}}=a^\mu b^\nu+a^\nu b^\mu$.
The four-vector $S_{LT}^\mu$ and tensor $S_{TT}^{\mu\nu}$ refer to the longitudinal-transverse and transverse-transverse tensor polarization.
The corresponding systematic definitions and physical interpretations have been discussed in ref.~\cite{Bacchetta:2000jk}.

As described by the Trento conventions, one can define two normalized vectors $\hat{t}$ and $\hat{z}$ as
\begin{align}
    \hat{t}^{\mu}  =\frac{2x_d}{\sqrt{1+\gamma^{2}}Q}\left(P^{\mu}-\frac{P\cdot q}{q^{2}}q^{\mu}\right),
    \quad
	\hat{z}^{\mu}  =-\frac{q^{\mu}}{Q}.\label{e.tzmu}
\end{align}
Since the lepton momenta lie in the $\hat{x}$-$\hat{z}$ plane, we can choose the transverse spatial basis vectors as 
\begin{align}
	\hat{x}^\mu =\frac{l_\perp^\mu} { \left| \bm{l}_\perp \right|}, \quad \hat{y}^\mu=\epsilon_\perp^{\mu\nu}\hat{x}_\nu.	\label{e.xymu}
\end{align}
Following the definitions of the light-cone basis vectors
\begin{align}
	\bar{n}^\mu=\frac{1}{\sqrt{2}}(\hat{t}^\mu+\hat{z}^\mu),\quad n^\mu=\frac{1}{\sqrt{2}}(\hat{t}^\mu-\hat{z}^\mu),\label{e.nnbar}
\end{align}
one can obtain the expression of the spin tensors in terms of the external momenta of the reaction with the basis vectors replaced by eqs.~\eqref{e.tzmu}--\eqref{e.xymu}.
The covariant form of the spin vector $S$ is given by
\begin{align}
    S^\mu=S_L \frac{P^\mu-q^\mu M^2/(P\cdot q)}{M\sqrt{1+\gamma^2}}+S_T^\mu,
\end{align}
where the longitudinal and transverse components are expressed as
\begin{align}
    S_L=\frac{S\cdot q}{P\cdot q}\frac{M}{\sqrt{1+\gamma^2}},
    \qquad 
    S_T^\mu=g_\perp^{\mu\nu}S_\nu.\label{e.SLST}
\end{align}
The azimuthal angle $\phi_T$, measured from the lepton plane to the transverse spin vector, is defined analogous to $\phi_h$ in eq.~\eqref{e.phih}, with $P_h$ replaced by $S$.
We can also express the spin tensor $T^{\mu\nu}$ in terms of the kinematic variables as
\begin{align}
    T^{\mu\nu}=\frac{1}{2}\bigg\{
    \frac{2}{3}S_{LL}\Big(\frac{2(\gamma^2 q^\mu- 2 x_d P^\mu)(\gamma^2 q^\nu-2 x_d P^\nu)}{\gamma^2(\gamma^2+1)Q^2} + g_\perp^{\mu\nu}\Big)
    +\frac{2x_d P^{\{\mu}S_{LT}^{\nu\}} -\gamma^2 q^{\{\mu}S_{LT}^{\nu\}}} {\gamma\sqrt{\gamma^2+1}Q}
    +S_{TT}^{\mu\nu}
    \bigg\},
\end{align}
where the expression of $g_\perp^{\mu\nu}$ is given by eq.~\eqref{e.gperp}.
Similar to the definition of eq.~\eqref{e.SLST}, the spin components of the spin tensor can be written as 
\begin{align}
    S_{LL}=&\frac{3}{2}T^{\mu\nu}\hat{L}_\mu\hat{L}_\nu,\\
    S_{LT}^\mu=&-2T_{\alpha\beta}\hat{L}^\alpha g_\perp^{\mu\beta},\\
    S_{TT}^{\mu\nu}=&2T_{\alpha\beta}g_\perp^{\alpha\mu}g_\perp^{\beta\nu} - T_{\alpha\beta}\hat{L}^\alpha\hat{L}^\beta g_\perp^{\mu\nu},
\end{align}
where $L^{\mu}$ is a longitudinal basis vector,
\begin{align}
  \hat{L}^\mu=-\frac{M}{2P\cdot n}n^\mu+\frac{P\cdot n}{M}\bar{n}^\mu.
\end{align}
The azimuthal angle associated with the tensor polarization can be defined as Lorentz covariant forms,
\begin{align}
    &\cos\phi_{LT}=-\frac{2T^{\mu\nu}\hat{L}_{\mu}\hat{x}_{\nu}}{|S_{LT}|},\quad
    \sin\phi_{LT}=-\frac{2T^{\mu\nu}\hat{L}_{\mu}\hat{y}_{\nu}}{|S_{LT}|},\\
    &\cos 2\phi_{TT}=\frac{2T^{\mu\nu}\hat{x}_{\mu}\hat{x}_{\nu}+\frac{2}{3}T^{\mu\nu}\hat{L}_{\mu}\hat{L}_{\nu}}{|S_{TT}|},
    \quad
    \sin 2\phi_{TT}=\frac{2T^{\mu\nu}\hat{x}_\mu\hat{y}_\nu}{|S_{TT}|},
\end{align}
where $|S_{LT}|$ and $|S_{TT}|$ are normalized factors, defined by
\begin{align}
    |S_{LT}|=\sqrt{-S_{LT}^\mu S_{LT\mu}},
    \quad
    |S_{TT}|=\sqrt{S_{TT}^{\mu\nu} S_{TT\mu\nu}}.
\end{align}
Since the quantities defined above are Lorentz invariant scalars, one can in principle evaluate them in any reference frame, although the meaning of the transverse or longitudinal direction is more clearly understood in a specific frame.

\section{Cross section in terms of structure functions}
\label{s.crosssection}
With one-photon-exchange approximation, the differential cross section of the SIDIS process can be expressed as the contraction of the leptonic tensor and the hadronic tensor~\cite{Bacchetta:2006tn},
\begin{align}
	\frac{d\sigma}{dx_d dy dz d\phi_h d\psi d P_{h\perp}^2}=\frac{\alpha^2 y}{8Q^4 z}L_{\mu\nu}W^{\mu\nu},\label{e.dsigma}
\end{align}
where $\alpha$ is the electromagnetic fine structure constant and the angle $\psi$ is the azimuthal angle of the outgoing lepton around the lepton beam axis, measured relative to an arbitrary fixed reference direction.
The leptonic tensor $L_{\mu\nu}$ is given by
\begin{align}
L_{\mu\nu}=2\left(l_{\mu}l_{\nu}^{\prime}+l_{\nu}l_{\mu}^{\prime}-g_{\mu\nu} l\cdot l^{\prime} + i\lambda_{e}\epsilon_{\mu\nu\rho\sigma}l^{\rho}l^{\prime\sigma}\right),
\end{align}
where $\lambda_e$ denotes the helicity of the lepton beam. 
The hadronic tensor $W^{\mu\nu}$ is expressed as
\begin{align}
	W^{\mu\nu}\left( q;P,S,T; P_h\right)=
    &\sum_{X}\delta^4\left(P+q-P_h-P_X\right)  \left\langle P,S,T | J^\mu(0) | P_X, P_h \right\rangle 
	\left\langle P_X, P_h| J^\nu(0) |P,S,T\right\rangle,\label{e.hadronictensor}
\end{align}
where $J^\mu$ is the electromagnetic current operator, and the sum $\displaystyle{\sum_{X}}$ runs over the final-state hadronic states $X$ with the momentum integration implicitly assumed. 
The hadronic tensor fulfills the constraints imposed by the Hermiticity and parity invariance,
\begin{align}
	W^{* \mu \nu}\left( q;P,S,T; P_h\right)&=W^{\nu \mu}\left( q;P,S,T; P_h\right), \label{e.Hermiticity}\\
	W^{\mu \nu}\left( q;P,S,T; P_h\right)&=W_{\mu \nu}\left(q ,P,-\bar{S},\bar{T}; P_h\right),\label{e.parity}
\end{align}
where $\bar{S}$ and $\bar{T}$ indicate a sign flip of all space components. 
Additionally, the gauge invariance requires 
\begin{align}
    q_\mu W^{\mu\nu}\left( q;P,S,T; P_h\right)=W^{\mu\nu} \left( q;P,S,T; P_h\right)q_\nu =0.\label{e.gaugeinvariace}
\end{align}
To impose this condition, one can construct the so-called conserved vectors and tensors using the four momenta $P$ and $q$,
\begin{align}
	P_q^\mu&=P^\mu-\frac{P\cdot q}{q^2}q^\mu,\label{e.pqmu}\\
	P_{hq}^\mu&=P_h^\mu-\frac{P_h\cdot q}{q^2}q^\mu,\\
    g^{\mu\nu}_q &= g^{\mu\nu} - \frac{q^\mu q^\nu}{q^2},\label{e.gqmunu}
\end{align}
which will vanish upon contraction with $q$.

Through kinematic analysis, the hadronic tensor can be typically expressed as a sum of basis Lorentz tensors multiplied by scalar functions, referred to as structure functions.
The basis Lorentz tensors are constructed from the available kinematic variables involved in this process, $P^\mu$, $P_h^\mu$, $q^\mu$, $S^\mu$, and $T^{\mu\nu}$.
While we can, in principle, construct the basis Lorentz tensors by exhaustively listing all possible combinations of the conserved vectors and tensors, this method quickly loses its practicality as the number of the measured momenta increases.
Instead of directly building all basis Lorentz tensors, we adopt a systematic procedure presented in refs.~\cite{Chen:2016moq,Jiao:2022gzu,Zhao:2022lbw,Zhao:2024zpy}.
Using the conserved vectors and tensors~\eqref{e.pqmu}--\eqref{e.gqmunu}, one can first obtain nine basic Lorentz tensors~\cite{Jiao:2022gzu,Zhao:2024zpy},
\begin{align}
	h_{U}^{S\mu\nu} & =\left\{ g^{\mu\nu}_q,P_{q}^{\mu}P_{q}^{\nu},P_{q}^{\{\mu}P_{hq}^{\nu\}},P_{hq}^{\mu}P_{hq}^{\nu}\right\} ,\label{e.hus}\\
	\tilde{h}_{U}^{S\mu\nu} & =\left\{ \epsilon^{\left\{ \mu qPP_{h}\right.}P_{q}^{\nu\}}, \epsilon^{\left\{ \mu qPP_{h}\right.}P_{hq}^{\nu\}}\right\}, \label{e.tildehus}\\
	h_{U}^{A\mu\nu} & =\left\{ P_{q}^{[\mu}P_{hq}^{\nu]}\right\}, \label{e.hua}\\
	\tilde{h}_{U}^{A\mu\nu} & =\left\{ \epsilon^{\mu\nu qP},\epsilon^{\mu\nu qP_{h}}\right\},\label{e.tildehua}
\end{align}
where the subscript $U$ denotes the unpolarized part,
the superscripts $S$ and $A$ stand for the symmetric and antisymmetric tensors, respectively.
The parity-conserving and parity-violating terms are labeled by $h$ and $\tilde{h}$, respectively.
We introduce the shorthand notations to make the expressions more compact, such as $\epsilon^{\mu\nu qP}\equiv\epsilon^{\mu\nu\rho\sigma}q_\rho P_\sigma$.
Under the constraint of parity invariance~\eqref{e.parity}, the unpolarized basis tensors are directly provided by those in $h_{U}^{S\mu\nu}$ and $h_{U}^{A\mu\nu}$.
If we take the parity-violating channels into account, e.g., the $Z$ boson exchange, the tensors from $\tilde{h}_{U}^{S\mu\nu}$ and $\tilde{h}_{U}^{A\mu\nu}$ should also be included.

Since the four vectors $P^\mu$, $q^\mu$, $P_h^\mu$, and $\epsilon^{\mu qPP_{h}}$ form the complete basis of four-dimensional spacetime, the spin vector and spin tensor can be expressed as the combinations of these basis with spin-dependent scalar coefficients,
as demonstrated in ref.~\cite{Zhao:2024zpy}. 
Consequently, the polarized basis Lorentz tensors can be constructed from the spin-independent basic Lorentz tensors in eqs.~\eqref{e.hus}--\eqref{e.tildehua} multiplied by a spin-dependent scalar or pseudoscalar.  
Here, focusing on the tensor-polarized part of a spin-1 target, we list the 16 basis tensors,
\begin{align}
	h_{T}^{S\mu\nu} & =\left\{ T^{P_hP_h},T^{P_hq},T^{qq}\right\} h_{U}^{S\mu\nu}, \left\{ \epsilon^{T^{P_h}PP_{h}q},\epsilon^{T^{q}PP_{h}q}\right\} \tilde{h}_{U}^{S\mu\nu}, \label{e.hTS}
\end{align}
which contribute to the symmetric part of $W^{\mu\nu}$, and the seven ones 
\begin{align}
	h_{T}^{A\mu\nu} & =\left\{ T^{P_hP_h},T^{P_hq},T^{qq}\right\} h_{U}^{A\mu\nu}, \left\{ \epsilon^{T^{P_h}PP_{h}q}, \epsilon^{T^{q}PP_{h}q}\right\} \tilde{h}_{U}^{A\mu\nu}  \label{e.hTA}
\end{align}
contributing to the antisymmetric part of $W^{\mu\nu}$, where the subscript $T$ indicates the tensor-polarized components.

With the basis tensors defined above, the tensor-polarized dependent hadronic tensor can be expanded as
\begin{align}
    W_T^{\mu \nu}=&\sum_{i=1}^{16} V_{T, i}^S  h_{T, i}^{S\mu\nu} 
	+i\sum_{i=1}^{7} V_{T, i}^A h_{T, i}^{A\mu\nu},\label{e.wT}
\end{align}
where the coefficients $V_{i}$, known as structure functions, are scalar functions of $Q^2$, $P\cdot q$, $P_h\cdot q$, and $P\cdot P_h$. 
Note that the above procedure for constructing basis Lorentz tensors does not apply to the inclusive DIS process, as the number of momenta in the hadronic tensor is insufficient to form a complete basis of the spacetime.

After contracting the hadronic tensor and the leptonic tensor, one can obtain the general form of the differential cross section for SIDIS process with a spin-1 target. 
For clarity, we decompose the cross section into three parts, 
\begin{align}
    d\sigma=d\sigma_{U}+d\sigma_{V}+d\sigma_{T},
\end{align}
where $d\sigma_U$, $d\sigma_V$, and $d\sigma_T$ represent the contributions from the unpolarized, vector-polarized, and tensor-polarized components, respectively.
To emphasize the tensor-polarized dependent cross section, we denote it with the subscript ``$\rm Tens$" in the following.
According to the azimuthal modulations and the spin states of the target, the cross section can be expressed in terms of 41 structure functions.
Among them, the 23 tensor-polarized ones only exist when the target has spin $s\geq 1$, which is also included in ref.~\cite{Chen:2020ugq}, while the remaining 18 ones also exist for SIDIS with a spin-1/2 target, as defined in ref.~\cite{Bacchetta:2006tn}.
Here we give the explicit expression of the tensor-polarized dependent cross section as
\begin{align}
	&\frac{d\sigma_{\rm Tens}}{dx_d dydzd\phi_h d\psi dP_{h\perp}^2}\nonumber\\
    &= \frac{\alpha^{2}}{x_d yQ^{2}}\frac{y^{2}}{2(1-\varepsilon)} \left(1+\frac{\gamma^{2}}{2x_d}\right) 
    \bigg\{S_{LL} \bigg[F_{U(LL) ,T}
+ \varepsilon
F_{U(LL) ,L}
+ \sqrt{2\,\varepsilon (1+\varepsilon)}\,\cos\phi_h\,
F_{U(LL)}^{\cos\phi_h}\nonumber
\\  &\qquad
+ \varepsilon \cos(2\phi_h)\, 
F_{U(LL)}^{\cos 2\phi_h}
+ \lambda_e\, \sqrt{2\,\varepsilon (1-\varepsilon)}\, 
           \sin\phi_h\, 
F_{L(LL)}^{\sin\phi_h}
\bigg]\nonumber\\
%
&\qquad+|S_{LT}|\bigg[\cos{(\phi_h-\phi_{LT})} \Big(F_{U(LT),T}^{\cos{(\phi_h-\phi_{LT}})}+\varepsilon F_{U(LT),L}^{\cos{(\phi_h-\phi_{LT}})}\Big)\nonumber\\
&\qquad+\sqrt{2\varepsilon(1+\varepsilon)} \Big( \cos{\phi_{LT}} F_{U(LT)}^{\cos{\phi_{LT}}}+ \cos{(2\phi_h-\phi_{LT})} F_{U(LT)}^{\cos{(2\phi_h-\phi_{LT}})}\Big) \nonumber\\
&\qquad+\varepsilon \Big(  \cos(\phi_h+\phi_{LT}) F_{U(LT)}^{\cos{(\phi_h+\phi_{LT}})} +\cos(3\phi_h-\phi_{LT}) F_{U(LT)}^{\cos{(3\phi_h-\phi_{LT}})}\Big)\nonumber\\
&\qquad+\lambda_e \sqrt{2\varepsilon(1-\varepsilon)} \Big(\sin{\phi_{LT}} F_{L(LT)}^{\sin{\phi_{LT}}} +\sin{(2\phi_h-\phi_{LT})} F_{L(LT)}^{\sin{(2\phi_h-\phi_{LT}})}\Big)\nonumber\\
&\qquad+\lambda_e \sqrt{1-\varepsilon^2} \sin{(\phi_h-\phi_{LT})} F_{L(LT)}^{\sin{(\phi_h-\phi_{LT}})}
\bigg]\nonumber\\
%
&\qquad+|S_{TT}|\bigg[\cos{(2\phi_h-2\phi_{TT})} \Big(F_{U(TT),T}^{\cos{(2\phi_h-2\phi_{TT}})}+\varepsilon F_{U(TT),L}^{\cos{(2\phi_h-2\phi_{TT}})}\Big)\nonumber\\
&\qquad+\sqrt{2\varepsilon(1+\varepsilon)} \Big( \cos{(\phi_h-2\phi_{TT})} F_{U(TT)}^{\cos{(\phi_h-2\phi_{TT})}}+ \cos{(3\phi_h-2\phi_{TT})} F_{U(TT)}^{\cos{(3\phi_h-2\phi_{TT}})}\Big) \nonumber\\
&\qquad+\varepsilon \Big(  \cos(2\phi_{TT}) F_{U(TT)}^{\cos{(2\phi_{TT}})} +\cos(4\phi_h-2\phi_{TT}) F_{U(TT)}^{\cos{(4\phi_h-2\phi_{TT}})}\Big)\nonumber\\
&\qquad+\lambda_e \sqrt{2\varepsilon(1-\varepsilon)} \Big(\sin{(\phi_h-2\phi_{TT})} F_{L(TT)}^{\sin{(\phi_h-2\phi_{TT}})} +\sin{(3\phi_h-2\phi_{TT})} F_{L(TT)}^{\sin{(3\phi_h-2\phi_{TT}})}\Big)\nonumber\\
&\qquad+\lambda_e \sqrt{1-\varepsilon^2} \sin{(2\phi_h-2\phi_{TT})} F_{L(TT)}^{\sin{(2\phi_h-2\phi_{TT}})}
\bigg]\bigg\},\label{e.SFs}
\end{align}
where the first and second subscripts of $F$ stand for the spin states of the incoming lepton and the target, respectively, while the third subscript, when present, specifies the polarization of the virtual photon.
The structure functions $F$, as scalar functions of $x_d$, $Q^2$, $z$, and $P_{h\perp}^2$, are linear combinations of the coefficients $V_{T, i}$ defined in eq.~\eqref{e.wT}. 
Therefore, the number of independent structure functions is entirely determined by the number of basis tensors derived from the kinematic analysis.
We note that the polarization of the target is defined relative to the photon direction, whereas, in experiments, the polarization with respect to the lepton beam direction is a more natural choice.
The conversion between the two sets of spin states 
is straightforward and is detailed in refs.~\cite{Diehl:2005pc,Cosyn:2024drt}.
The ratio of the longitudinal and the transverse photon flux is defined by
\begin{align}
	\varepsilon=\frac{1-y-\frac{1}{4}\gamma^2y^2}{1-y+\frac{1}{2}y^2+\frac{1}{4}\gamma^2y^2},\label{e.ratio}
\end{align}
with which the depolarization factors in the cross section can be expressed in terms of $\gamma^2$ and $y$~\cite{Bacchetta:2006tn}.

After integrating eq.~\eqref{e.SFs} over the transverse momentum $P_{h\perp}$, one can find that the terms that depend on the azimuthal angle $\phi_h$ vanish.
Consequently, the tensor-polarized part of the cross section can be expressed in terms of the following five structure functions,
\begin{align}
    \frac{d\sigma_{\rm Tens}}{dx_d dy d\psi dz}  =&\frac{2\alpha^2}{x_dyQ^2}\frac{y^2}{2\left(1-\varepsilon\right)}\bigg(1+\frac{\gamma^2}{2x_d}\bigg)
    \bigg\{
        S_{LL} \bigg(F_{U(LL),T}+\varepsilon F_{U(LL),L}\bigg)\nonumber\\
     &+ |S_{LT}| \bigg(\sqrt{2\varepsilon(1+\varepsilon)} \cos\phi_{LT} F_{U(LT)}^{\cos\phi_{LT}}
     +\lambda_e \sqrt{2\varepsilon(1-\varepsilon)} \sin\phi_{LT} F_{L(LT)}^{\sin\phi_{LT}}\bigg)\nonumber\\
     &+ |S_{TT}| \varepsilon \cos(2\phi_{TT}) F_{U(TT)}^{\cos(2\phi_{TT})} 
     \bigg\},\label{e.dsigmadz}
\end{align}
where the integrated structure functions, no longer dependent on $P_{h\perp}$, are given by
\begin{align}
    F_{U(LL),T}(x_d,z,Q^2)=\int d^2\boldsymbol{P}_{h\perp}F_{U(LL),T}(x_d,z,P_{h\perp}^2,Q^2).
\end{align}
This relation is also applicable to the other structure functions in eq.~\eqref{e.dsigmadz}.
These integrated functions can be used to study the collinear PDFs or FFs.

The general differential cross section of the inclusive DIS off a tensor-polarized target can be obtained by integrating eq.~\eqref{e.dsigmadz} over the longitudinal momentum fraction $z$,
\begin{align}
    \sum_{h} \int d z z \frac{d \sigma(\ell d \rightarrow \ell h X)}{d z d x_d d y d \psi}= \frac{1}{\nu} \sum_{h} \int d E_{h}\,E_{h} \frac{d \sigma(\ell d \rightarrow \ell h X)}{d E_{h} d x_d d y d \psi} = \frac{\nu+M}{\nu} \frac{d \sigma(\ell d \rightarrow \ell X)}{d x_d d y d \psi},
    \end{align}
where $E_h=(P\cdot P_h)/M$ is the energy of the produced hadron in the target rest frame.
When summing over all hadrons in the final state, the total energy is given by $\nu+M$.
Factoring out the overall kinematic factor, $(\nu+M)/\nu = 1+\gamma^2 /(2x_d)$, originated from the integration over $z$ and the sum over hadrons, we have 
\begin{align}
    \frac{d\sigma_{\rm Tens}}{dx_d dy d\psi} =&\frac{2\alpha^2}{x_d yQ^2}\frac{y^2}{2\left(1-\varepsilon\right)}
    \bigg\{
        S_{LL} \bigg(F_{U(LL),T}+\varepsilon F_{U(LL),L}\bigg)\nonumber\\
     &+ |S_{LT}| \sqrt{2\varepsilon(1+\varepsilon)} \cos\phi_{LT} F_{U(LT)}^{\cos\phi_{LT}}
     %
     + |S_{TT}| \varepsilon \cos(2\phi_{TT}) F_{U(TT)}^{\cos(2\phi_{TT})} 
     \bigg\},\label{e.dsigmaDIS}
\end{align}
where the structure functions, as the functions of $x_d$ and $Q^2$, are defined for inclusive process, 
\begin{align}
    F_{U(LL),T}(x_d,Q^2)=\sum_h \int dz\,z\,F_{U(LL),T}(x_d,z,Q^2),\label{e.intdz}
\end{align}
with analogous definitions for the other structure functions.
We note that the time reversal invariance requires~\cite{Cosyn:2024drt}
\begin{align}
    \sum_h \int dz \, z \,F_{L(LT)}^{\sin\phi_{LT}}(x_d,z,Q^2) =0.
\end{align}

Historically, four tensor-polarized structure functions for inclusive DIS, denoted as $b_{1-4}$, were first introduced in 1989 in ref.~\cite{Hoodbhoy:1988am} by utilizing an alternative approach to decompose the hadronic tensor.  The relations between these two sets of structure functions were established in ref.~\cite{Cosyn:2024drt}. We report them here for convenience:
\begin{align}
F_{U (LL),T}   &=
-\frac{2}{3}\biggl[2\bigl(1+\gamma^2\bigr)b_1-\frac{\gamma^2}{x_d}
\biggl(\frac{1}{6} b_2-\frac{1}{2} b_3\biggr)\biggr ],\label{e.brelation1}
\\
F_{U (LL),L}   &=
\frac{2}{3x_d}\biggl[2\bigl(1+\gamma^2\bigr)x_d b_1
-(1+\gamma^2)^2
\biggl(\frac{1}{3}b_2+b_3+b_4\biggr)
\nonumber\\ & \quad 
-\bigl(1+\gamma^2\bigr)
\biggl(\frac{1}{3}b_2-b_4\biggr)
-\biggl(\frac{1}{3}b_2-b_3 \biggr)\biggr ],
\label{e.brelation2}
\\
F_{U(LT)}^{\cos\phi_{LT}}
  &=
-\frac{\gamma}{4x_d}\biggl[\bigl(1+\gamma^2\bigr)
\biggl(\frac{1}{3} b_2-b_4\biggr)
+\biggl(\frac{2}{3}b_2-2b_3\biggr)\biggr],\label{e.brelation3}
\\
F_{U(TT)}^{\cos(2\phi_{TT})} 
  &= - \frac{\gamma^2}{2x_d}
\biggl(\frac{1}{6}b_2-\frac{1}{2}b_3 \biggr).\label{e.brelation4}
\end{align}    
The structure functions $b_1$ and $b_2$ obey a Callan-Gross-like relation $b_2 =2 x_d b_1$ in the Bjorken limit. The structure functions defined in eq.~\eqref{e.dsigmaDIS} have advantages in illustrating the geometric dependence on the different tensor-polarized components and azimuthal modulations. We note that our conventions of the tensor polarization components differ from those used in ref.~\cite{Cosyn:2024drt}. They are related by $T_{LL} = 2 S_{LL} / 3$, $T_{LT}\cos\phi_{T_L} = S_{LT}^x / 2$, and $T_{TT}\cos2\phi_{T_T} = S_{TT}^{xx}$, where $T_{LL}$, $T_{LT}$, $T_{TT}$, $\phi_{T_L}$, and $\phi_{T_T}$ are the tensor polarization components and azimuthal angles used in ref.~\cite{Cosyn:2024drt}. These differences can be absorbed into the normalization of corresponding structure functions, and the formal expression of differential cross section keeps the same. Accordingly, the right-hand sides of eqs.~\eqref{e.brelation1} and \eqref{e.brelation2} differ by a factor of $2/3$ from eqs.~(17a) and (17b) in ref.~\cite{Cosyn:2024drt}, and the right-hand side of eq.~\eqref{e.brelation3} differs by a factor of $1/2$ from eq.~(17d) in ref.~\cite{Cosyn:2024drt}. The corresponding factors due to such normalization differences are also reflected in eqs.~\eqref{e.b1}-\eqref{e.b4}.

\section{Structure functions in terms of partonic functions}
\label{s.partonmodel}
\subsection{Hadronic tensor up to subleading order in $1/Q$}
Now we calculate the cross section in the parton model.
We find that it is convenient to discuss the distribution and fragmentation functions in the light-cone coordinates, where the hadron momenta have no transverse component, while the virtual photon is characterized by the transverse momentum $q_T$.
Thus, we choose the target and the detected hadron back-to-back frame where the momenta can be decomposed as
\begin{align}
    P^{\mu}=P^{+} n_{+}^{\mu}+\frac{M^{2}}{2P^{+}}n_{-}^{\mu},\quad 
    P_{h}^{\mu}=\frac{M_{h}^{2}}{2P_{h}^{-}} n_{+}^{\mu}+P_{h}^{-} n_{-}^{\mu}.
\end{align}
Here, we label the two light-cone vectors by $n_+$ and $n_-$ to distinguish them from the definitions of $\bar{n}$ and $n$ in eq.~\eqref{e.nnbar}.
They can be expressed in terms of $\hat{t}$, $\hat{z}$, and the transverse momentum $q_T$~\cite{Mulders:1995dh,Boer:2003cm}, 
\begin{align}
 n_+^\mu=\frac{1}{\sqrt{2}}(\hat{t}^\mu+\hat{z}^\mu),
 \quad
 n_-^\mu=\frac{1}{\sqrt{2}}(\hat{t}^\mu-\hat{z}^\mu-2\frac{q_T^\mu}{Q}).
\end{align}
The relation of the transverse vectors between the two coordinates systems can be derived from the transverse tensor
\begin{align}
    g_T^{\mu\nu}=g_\perp^{\mu\nu}+\frac{1}{Q}\hat{z}^{\{\mu}q_T^{\nu\}} +\frac{1}{Q}\hat{t}^{\{\mu}q_T^{\nu\}}.
\end{align}
The tensor $g_\perp^{\mu\nu}$, also referred to as perpendicular projection tensor, is defined by eq.~\eqref{e.gperp}.
Note that the choice of the coordinate system is different from that in section~\ref{s.kinematics}, where the transverse direction is defined relative to the momenta of the target and the virtual photon hadron, rather than the momenta of the target and the produced hadron. 
Further details on the relations between the two frames have been discussed in refs.~\cite{Mulders:1995dh,Boer:2003cm}.

\begin{figure}[htbp]
    \centering
    \begin{subfigure}[b]{0.35\textwidth}
        \includegraphics[width=\textwidth]{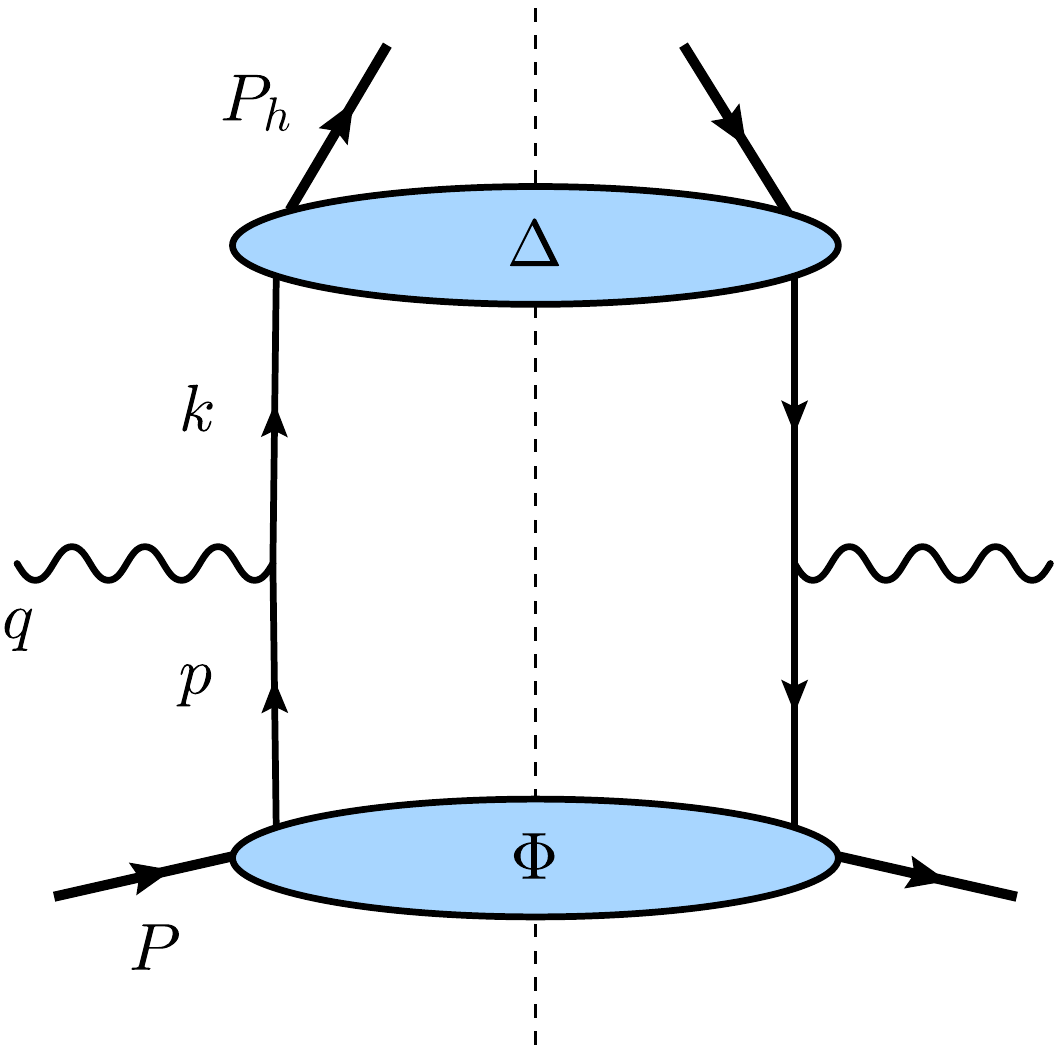}
        \caption{}
        \label{f.hta}
    \end{subfigure}
    \hfill
    \begin{subfigure}[b]{0.35\textwidth}
        \includegraphics[width=\textwidth]{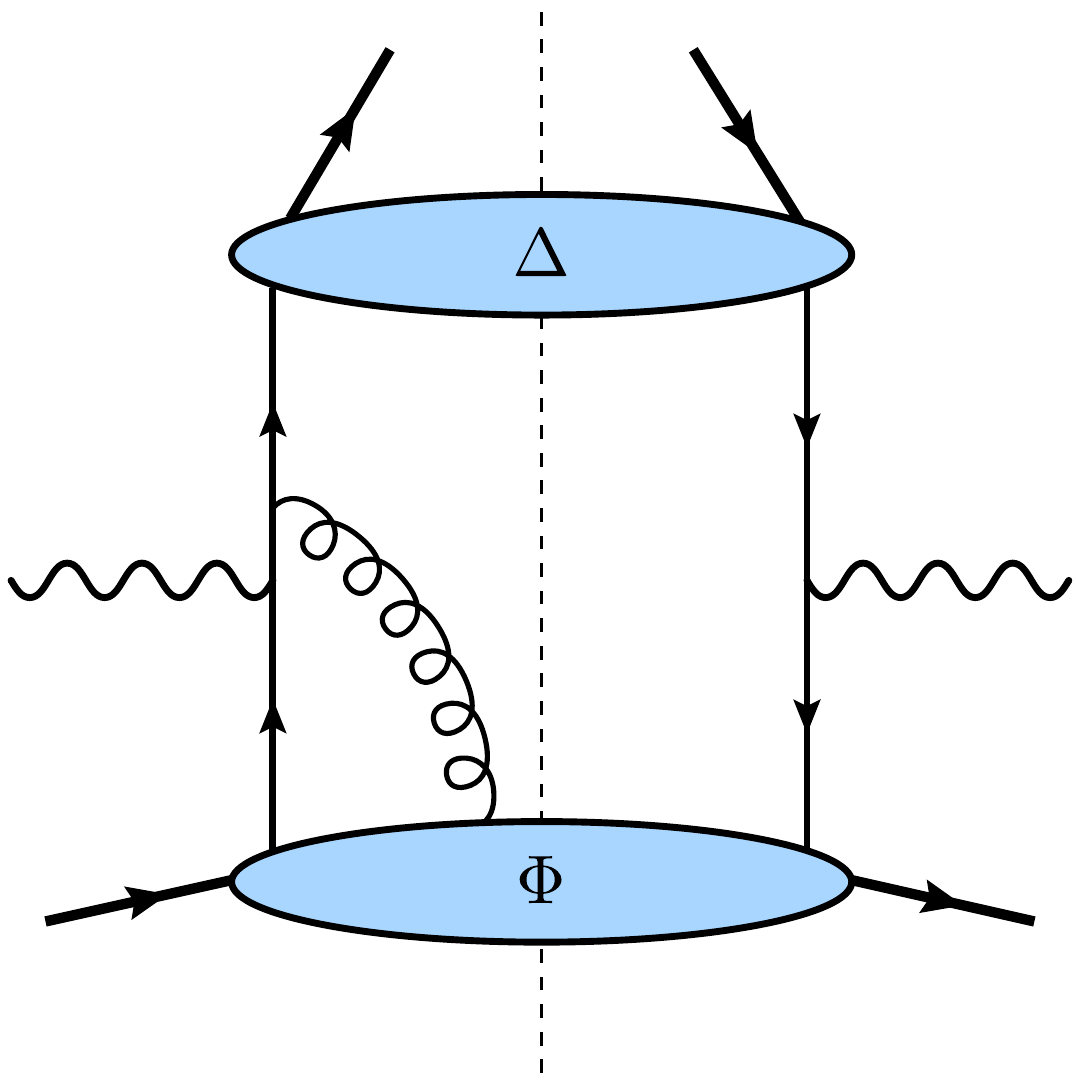}
        \caption{}
        \label{f.htb}
    \end{subfigure}
    
    \vspace{\baselineskip} 
    
    \begin{subfigure}[b]{0.35\textwidth}
        \includegraphics[width=\textwidth]{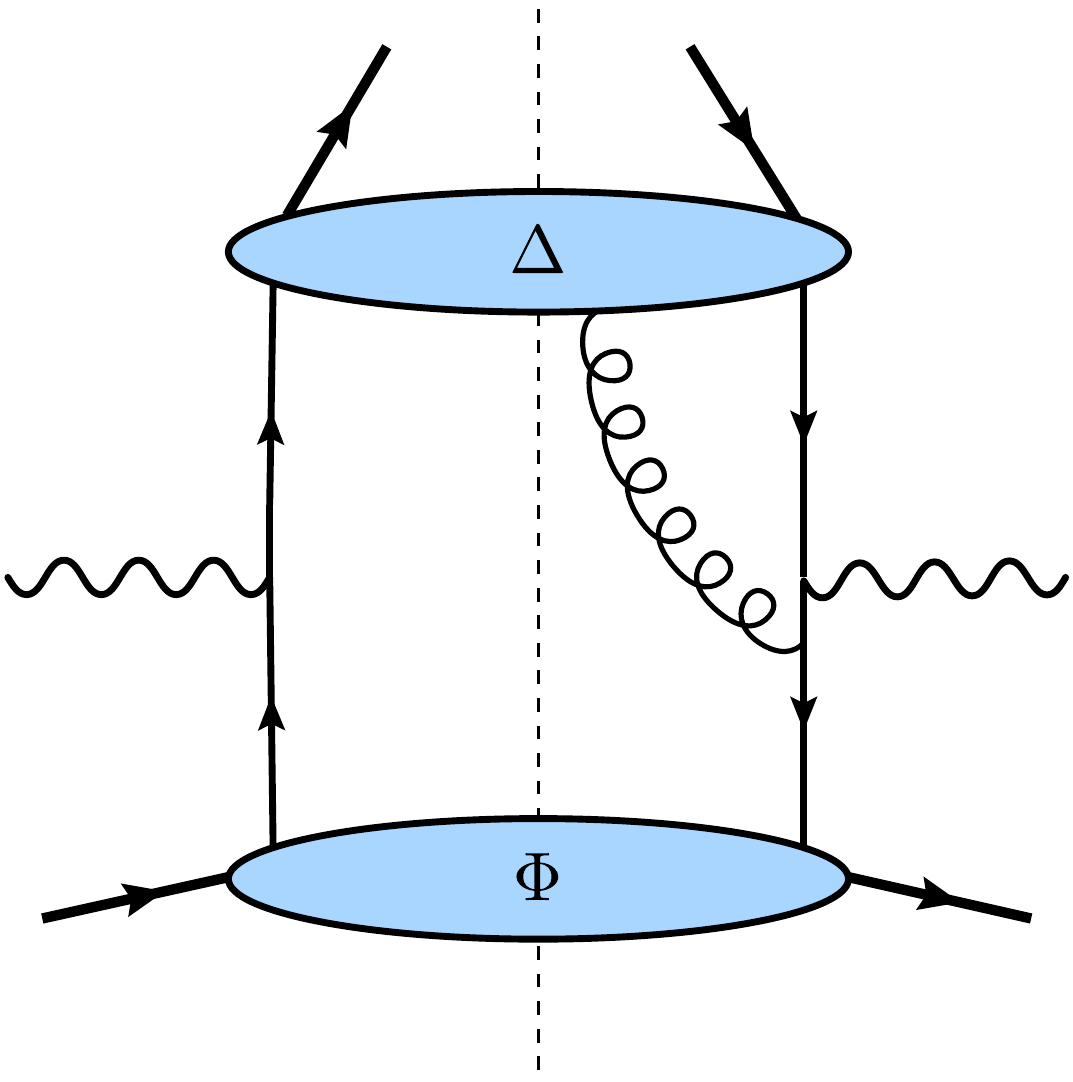}
        \caption{}
        \label{f.htc}
    \end{subfigure}
    \hfill
    \begin{subfigure}[b]{0.35\textwidth}
        \includegraphics[width=\textwidth]{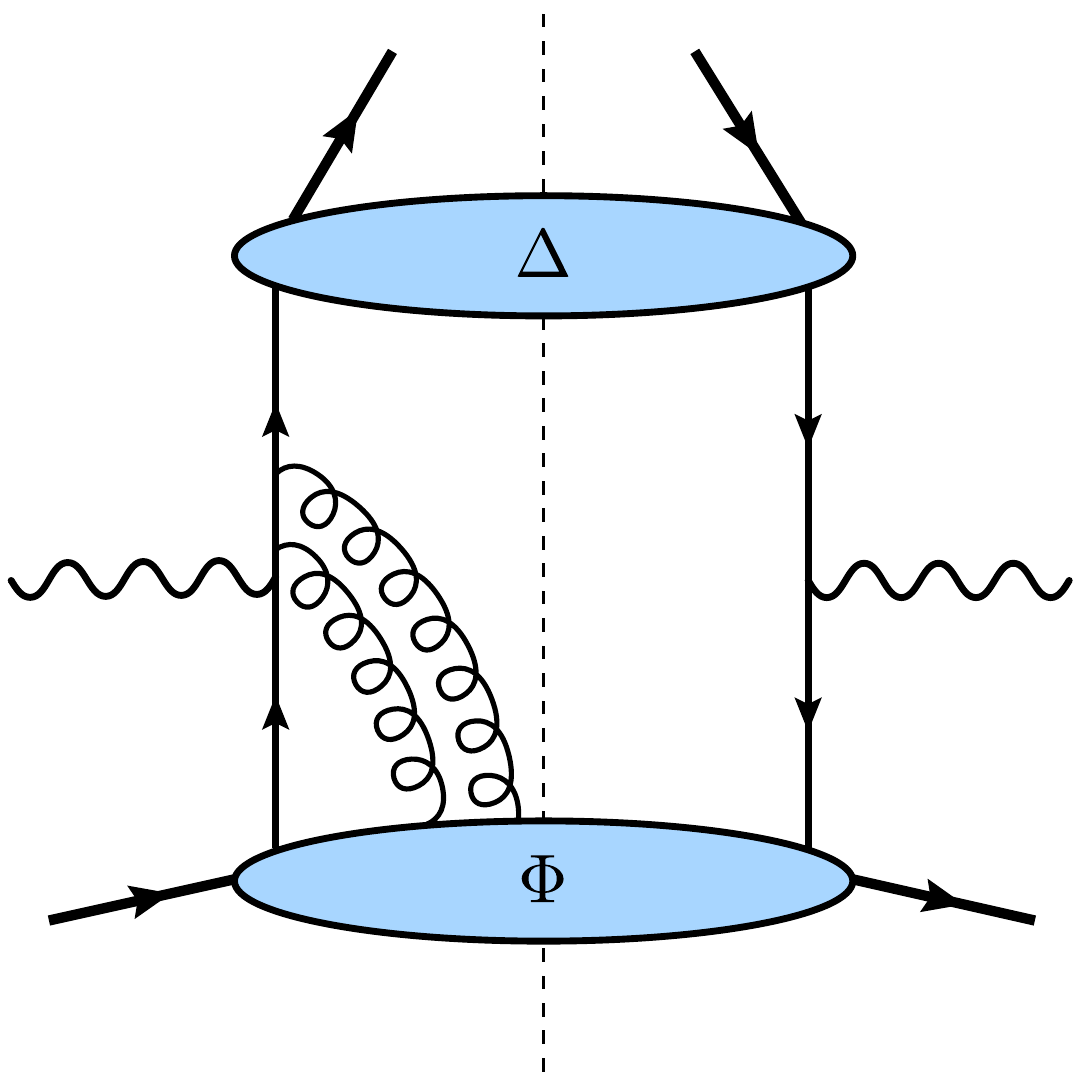}
        \caption{}
        \label{f.htd}
    \end{subfigure}
    \caption{Illustration of the hadronic tensor in the $1/Q$ expansion for SIDIS.}
    \label{f.hadrontensor}
\end{figure}

We consider the SIDIS process in the kinematic region in which the target and the produced hadron are back-to-back with the squared transverse momentum of photon $\bm{q}_T^2$ much smaller than the hard scale $Q^2$, i.e., $\bm{q}_T^2 \ll Q^2$. 
In this limit, one can apply the TMD factorization in which the hadronic tensor is expressed as a hard photon-quark scattering process convoluting with the nonperturbative functions, namely, the TMD PDFs and TMD FFs. We restrict the calculation to the leading and first subleading terms in the $1/Q$ expansion of the cross section and only consider the graphs with hard scattering at the tree level. 
Up to $\mathcal{O}(1/Q)$, we need to include the handbag diagrams with an additional gluon originating from the correlator associated with either the target or the produced hadron. They are still to be considered as tree-level diagrams. A few remarks on the effect of higher-order diagrams will be mentioned in section~\ref{s.strucfuncresults}.

The hadronic tensor can then be obtained from the diagrams illustrated in figure~\ref{f.hadrontensor}.
Its corresponding expression is given by \cite{Mulders:1995dh,Boer:2003cm}
\begin{align}
   &W^{\mu\nu}=2z\sum_{a}e_{a}^{2}\int d^{2}\boldsymbol{p}_{T}d^{2}\boldsymbol{k}_{T}\delta^{2}(\boldsymbol{p}_{T}+\boldsymbol{q}_{T}-\boldsymbol{k}_{T})\mathrm{Tr}\bigg\{\Phi^{a}(x,p_{T})\gamma^{\mu}\Delta^{a}(z,k_{T})\gamma^{\nu}\nonumber\\
  &-\frac{1}{Q\sqrt{2}}\bigg[\gamma^{\alpha}\slashed{n}_+\gamma^{\nu}\tilde{\Phi}_{A\alpha}^{a}(x,p_{T})\gamma^{\mu}\Delta^{a}(z,k_{T})
  +\gamma^{\alpha}\slashed{n}_-\gamma^{\mu}\tilde{\Delta}_{A\alpha}^{a}(z,k_{T})\gamma^{\nu}\Phi^{a}(x,p_{T})+\mathrm{h.c.}\bigg]\bigg\},\label{e.subwmunu} 
\end{align}
with the light-cone longitudinal momentum fraction defined by $xP^+=P_h^-/z=Q/\sqrt{2}$.
Here $x$ equals the scaling variable $x_d$ at the leading order when neglecting the target mass.
The $p_T$ and $k_T$ correspond to the transverse momenta of the quark distributing in the target and the quark fragmenting into the observed hadron, respectively.
The sum over “$a$” runs over all active quark and antiquark flavors, and $e_a$ denotes the charge in units of the positron charge.
The terms with $n_+$ and $n_-$ in eq.~\eqref{e.subwmunu} arise from fermion propagators in the quark-photon hard scattering part while neglecting the suppression by $Q^2$.
The explicit forms of the quark distribution correlator $\Phi$, the quark fragmentation correlator $\Delta$, and the quark-gluon-quark correlators $\tilde{\Phi}_A$ and $\tilde{\Delta}_A$ with an additional gluon leg will be provided in the following subsections. 
In eq.~\eqref{e.subwmunu}, the first, second, and third terms correspond to the diagrams shown in figures~\ref{f.hta},~\ref{f.htb}, and~\ref{f.htc}, where the gluon is transverse polarized. 
The h.c. in eq.~\eqref{e.hadronictensor} stands for the Hermitian conjugate of figures~\ref{f.htb} and~\ref{f.htc}, with the gluon on the opposite side of the final-state cut.

\subsection{Quark-quark correlators}
Before writing down the explicit expression of the hadronic tensor in the parton model, we first provide a brief review of the definitions of the tensor-polarized TMD PDFs for spin-1 hadrons. 

The quark-quark distribution correlation function is defined as 
\begin{align}
	\Phi_{ij}(x,p_T)&=\int\frac{d\xi^- d^2\bm{\xi}_T}{(2\pi)^3} e^{ip\cdot \xi} 
    \langle P| \bar{\psi}_j(0) 
    \mathcal{U}_{(0,\xi)}^{c}
    \psi_i(\xi)|P\rangle\bigg|_{\xi^+=0},
    \label{e.Phi}
\end{align} 
where $p^+=xP^+$ with $P^+$ being the large component of the target momentum. 
The subscripts $i$ and $j$ are Dirac indices, which are suppressed in the rest of the paper.
The gauge link $\mathcal{U}_{(0,\xi)}^{c}$, also called the Wilson line, is defined by a path integral
\begin{align}
    \mathcal{U}_{(0,\xi)}^{c}=\mathcal{P}\exp\bigg[-ig\int_{0}^{\xi} d\xi \cdot A(\xi)\bigg],\label{e.pathint}
\end{align}
which renders the correlation function gauge invariant. 
Here the superscript $c$ indicates the path of integral, $g$ represents the strong coupling constant, and $A(\xi)$ is the gluon field operator.
The antiquark correlation function can be defined by a similar expression with the quark field replaced by its charge conjugation.

The gauge link in the SIDIS process arises from the exchange of gluons between the hard scattering and the non-perturbative functions, as shown in figure~\ref{f.hadrontensor}.
The explanation of the gauge link in the SIDIS process has been discussed in refs.~\cite{Ji:2002aa,Belitsky:2002sm} and the detailed derivation for the leading terms in the $1/Q$ expansion is shown in refs.~\cite{Boer:2003cm,Collins:1981uw} .

Since the correlation function is a $4\times 4$ matrix in the Dirac space, it can be decomposed by a basis of 16 Dirac matrices.
By taking $P^+$ as the dominant momentum component of the hadron, the elements in the correlation functions are ordered in powers of $M/P^+$. Each additional power of $M/P^+$ will lead to a power of $M/Q$ suppressed contribution to the cross section~\cite{Levelt:1993ac}. Here the twist $t$ of TMD PDFs or FFs is associated to $(M/Q)^{t-2}$ in structure functions~\cite{Mulders:1995dh}.
The leading-twist (twist-2) PDFs can be projected from the correlation function using the Dirac matrices $\gamma^+$, $\gamma^+ \gamma_5$, and $i\sigma^{i+}\gamma_5$, while the twist-3 PDFs can be projected by $\gamma^i$, $\bm{1}$, $i\gamma_5$, $i\sigma^{ij}\gamma_5$, and $i\sigma^{-+}\gamma_5$.
For a spin-1 hadron, the quark-quark correlation function in terms of the TMD PDFs contains the unpolarized, vector-polarized, and tensor-polarized terms.
Among them, the unpolarized and vector-polarized parts are identical to those defined for the spin-1/2 hadron.
The tensor-polarized part of the quark-quark correlation function, denoted as $\Phi(x,p_{T};T)$, was first parametrized in terms of ten twist-2 TMD PDFs in ref.~\cite{Bacchetta:2000jk}, and subsequently decomposed into 20 twist-3 TMD PDFs, and ten twist-4 TMD PDFs, as shown in ref.~\cite{Kumano:2020ijt}.
Up to the subleading twist, a complete decomposition of the tensor-polarized dependent correlation function for a spin-1 hadron is given by
\begin{align}
    \Phi(x,p_{T};T)=&\frac{1}{2}\bigg\{
     f_{1LL}S_{LL}\slashed{n}_+
     -f_{1LT}\frac{S_{LT}\cdot p_{T}}{M}\slashed{n}_+
     +f_{1TT}\frac{p_{T}\cdot S_{TT}\cdot p_{T}}{M^2}\slashed{n}_+ \nonumber\\
    &+\Big(g_{1LT}\frac{\epsilon_T^{\mu\nu}S_{LT\mu} p_{T\nu}}{M}\gamma_5\slashed{n}_+\Big)
    -\Big(g_{1TT}\frac{\epsilon_T^{\mu\nu} S_{TT\nu\rho}p_T^\rho  p_{T\mu}}{M^2}\gamma_5\slashed{n}_+ \Big)\nonumber\\
    & +\Big(i h_{1LL}^\perp S_{LL}\frac{\big[\slashed{p}_T,\slashed{n}_+\big]}{2M}\Big)
     +\Big(i h_{1LT}^\prime \frac{\big[\slashed{S}_{LT},\slashed{n}_+\big]}{2}\Big)
     -\Big(i h_{1LT}^\perp \frac{S_{LT}\cdot p_T}{M} \frac{\big[\slashed{p}_T,\slashed{n}_+\big]}{2M}\Big) \notag\\
     &- \Big(h_{1TT}^\prime \frac{\sigma^{\mu\nu} S_{TT\mu\rho}p_T^\rho n_{+\nu}}{M}\Big)
     +\Big(i h_{1TT}^\perp\frac{p_{T}\cdot S_{TT}\cdot p_{T}}{M^2} \frac{\big[\slashed{p}_T,\slashed{n}_+\big]}{2M} \Big)\bigg\}\nonumber
     \\
     %
    &+\frac{M}{2P^{+}}\bigg\{
    e_{LL} S_{LL}-e_{LT}^\perp \frac{S_{LT}\cdot p_{T}}{M}
    +e_{TT}^\perp \frac{p_{T}\cdot S_{TT}\cdot p_{T}}{M^{2}}\nonumber\\
    &\quad+f_{LL}^{\perp} S_{LL}\frac{\slashed{p}_{T}}{M}+f_{LT}^{\prime} \slashed{S}_{LT}-f_{LT}^{\perp} \frac{\slashed{p}_{T}S_{LT}\cdot p_{T}}{M^{2}}\nonumber\\
    &\quad-f_{TT}^{\prime} \frac{S_{TT}^{\mu\nu}\gamma_{\mu}p_{T\nu}}{M}+f_{TT}^{\perp} \frac{p_{T}\cdot S_{TT}\cdot p_{T}}{M^{2}}\frac{\slashed{p}_{T}}{M}\nonumber\\
    &\quad-i e_{LT} \frac{S_{LT\mu}\epsilon_{T}^{\mu\nu}p_{T\nu}}{M}\gamma_{5}
    +i e_{TT} \frac{S_{TT\mu\rho}p_{T}^{\rho}\epsilon_{T}^{\mu\nu}p_{T\nu}}{M^{2}}\gamma_{5}\nonumber\\
    &\quad-\Big(g_{LL}^{\perp} S_{LL}\gamma_{5}\frac{\varepsilon_{T}^{\mu\nu}\gamma_{\mu}p_{T\nu}}{M}\Big)
    -\Big(g_{LT}^{\prime} \gamma_{5}\epsilon_{T}^{\mu\nu}\gamma_{\mu}S_{LT\nu}\Big)
    +\Big(g_{LT}^{\perp} \gamma_{5}\frac{\varepsilon_{T}^{\mu\nu}\gamma_{\mu}p_{T\nu}S_{LT}\cdot p_{T}}{M^{2}}\Big)\nonumber\\
    &\quad +\Big(g_{TT}^{\prime} \gamma_{5}\frac{\varepsilon_{T}^{\mu\nu}\gamma_{\mu}S_{TT\nu\alpha}p_{T}^{\alpha}}{M}\Big)
    -\Big(g_{TT}^{\perp} \frac{p_{T}\cdot S_{TT}\cdot p_{T}}{M^{2}}\gamma_{5}\frac{\varepsilon_{T}^{\mu\nu}\gamma_{\mu}p_{T\nu}}{M}\Big)\nonumber\\
    &\quad-\Big(ih_{LL}S_{LL}\frac{[\slashed{n}_-,\slashed{n}_+]}{2}\Big)
    -\Big(ih_{LT} \frac{S_{LT}\cdot p_{T}}{M}\frac{[\slashed{n}_-,\slashed{n}_+]}{2}\Big)
    +\Big(ih_{LT}^{\perp} \frac{[\slashed{S}_{LT},\slashed{p}_{T}]}{2M}\Big)
    \nonumber\\
    &\quad+\Big(ih_{TT} \frac{p_{T}\cdot S_{TT}\cdot p_{T}}{M^{2}}\frac{[\slashed{n}_-,\slashed{n}_+]}{2}\Big)
    -\Big(ih_{TT}^{\perp} \frac{[S_{TT}^{\mu\nu}\gamma_{\mu}p_{T\nu},\slashed{p}_{T}]}{2M^{2}}\Big)
    \bigg\},\label{e.TMDPDFs}
\end{align}
where we have $S_{LT}\cdot p_{T}=-\bm{S}_{LT}\cdot \bm{p}_{T}$, $p_{T}\cdot S_{TT}\cdot p_{T}=S_{TT}^{\mu\nu}p_{T\mu}p_{T\nu}$, and
$\epsilon_T^{\mu\nu}=\epsilon^{\mu\nu\rho\sigma} n_{+\rho} n_{-\sigma}$.
For conciseness, we have omitted the arguments ($x$, $p_T^2$) of the TMD PDFs. The TMD PDFs with a prime will be explicitly addressed later.
The functions in eq.~\eqref{e.TMDPDFs} with the order of $\mathcal{O}(1)$ correspond to the twist-2 TMD PDFs, while the terms proportional to $1/P^+$ represent twist-3 TMD PDFs.
As this paper focuses on twist-2 and twist-3 TMD PDFs, the twist-4 functions are not included here and can be found in ref.~\cite{Kumano:2020ijt}.
The TMD PDFs labeled by $f$ and $g$ are chiral-even functions, while those labeled by $h$ are chiral-odd functions.    
Note that the terms in parentheses are commonly referred to as the naively T-odd functions, which change sign under naive time-reversal transformation in the absence of initial- and final-state interactions. The time-reversal condition results in a minus sign in the gauge link, which cancels the sign generated by T-odd functions, thereby ensuring the time-reversal invariance of QCD~\cite{Ellis:1982wd,Collins:1981uw,Ji:2002aa,Belitsky:2002sm,Boer:2003cm}.
We adopt the notation conventions for spin-1 hadron TMD PDFs from ref.~\cite{Bacchetta:2000jk} for twist-2 contributions and ref.~\cite{Kumano:2020ijt} for twist-3 contributions, except for the twist-3 $h_{LL}$ term, whose sign in eq.~\eqref{e.TMDPDFs} is opposite to that in ref.~\cite{Kumano:2020ijt} to align with the sign of $h$.

We can collect the TMD PDFs defined in eq.~\eqref{e.TMDPDFs} through the projection by the Dirac matrices
\begin{align}
    \Phi^{[\Gamma]}(x,p_T;T)=\frac{1}{2}\textrm{Tr}[\Phi(x,p_T;T)\Gamma],
\end{align}
where $\Gamma$ is taken as $\gamma^+$, $\gamma^+ \gamma_5$, and $i\sigma^{i+}\gamma_5$ at leading twist, and $\Gamma$ equals $\gamma^i$, $\bm{1}$, $i\gamma_5$, $\gamma^i \gamma_5$, $i\sigma^{-+}\gamma_5$, and $i\sigma^{ij}\gamma_5$ at subleading twist.    
The expressions of the traces of the correlator $\Phi(x,p_T^2;T)$ are listed below: 
\begin{align}
\Phi^{[\gamma^{+}]}(x,p_{T};T)=& f_{1LL}S_{LL}-f_{1LT}\frac{S_{LT}\cdot p_{T}}{M}  +f_{1TT}\frac{p_{T}\cdot S_{TT}\cdot p_{T}}{M^{2}}, \label{e.phigamma}\\
\Phi^{[\gamma^{+}\gamma_{5}]}(x,p_{T};T)=& g_{1LT}\frac{S_{LT\mu}\varepsilon_T^{\mu\nu}p_{T\nu}}{M} +g_{1TT}\frac{S_{TT\mu\rho}p_{T}^{\rho}\varepsilon_{T}^{\mu\nu}p_{T\nu}}{M^{2}},\\
\Phi^{[i\sigma^{i+}\gamma_5]}(x,p_{T};T)=& h_{1LL}^{\perp}S_{LL} \frac{\epsilon_T^{ij}p_{Tj}}{M}  
+h^\prime_{1LT} \epsilon_T^{ij}S_{LTj}
-h_{1LT}^{\perp}\frac{S_{LT}\cdot p_T}{M} \frac{\epsilon_T^{ij}p_{Tj}}{M}
\nonumber\\ 
&-h^\prime_{1TT}\frac{\epsilon_T^{ij}S_{TTjl}p_T^l}{M}
+h_{1TT}^{\perp}\frac{p_T\cdot S_{TT}\cdot p_T}{M^2} \frac{\epsilon_T^{ij} p_{Tj}}{M},\label{e.sigmaip}\\
\Phi^{[\gamma^{i}]}(x,p_{T};T)=&\frac{M}{P^{+}}\bigg[f_{LL}^{\perp} S_{LL}\frac{p_{T}^{i}}{M}+f_{LT}^{\prime} S_{LT}^{i}-f_{LT}^{\perp} \frac{p_{T}^{i}S_{LT}\cdot p_{T}}{M^{2}}
-f_{TT}^{\prime} \frac{S_{TT}^{ij}p_{Tj}}{M}\nonumber\\
 & \qquad+f_{TT}^{\perp} \frac{p_{T}\cdot S_{TT}\cdot p_{T}}{M^{2}}\frac{p_{T}^{i}}{M}\bigg],\\
\Phi^{[1]}(x,p_{T};T)=&\frac{M}{P^{+}} \bigg[e_{LL} S_{LL}
-e_{LT}^\perp \frac{S_{LT}\cdot p_{T}}{M}
+e_{TT}^\perp \frac{p_{T}\cdot S_{TT}\cdot p_{T}}{M^{2}} \bigg],\\
\Phi^{[i\gamma_{5}]}(x,p_{T};T)=&\frac{M}{P^{+}} \bigg[
e_{LT} \frac{S_{LT\mu}\epsilon_{T}^{\mu\nu}p_{T\nu}}{M}
-e_{TT} \frac{S_{TT\mu\rho}p_{T}^{\rho}\epsilon_{T}^{\mu\nu}p_{T\nu}}{M^{2}} \bigg],\\
\Phi^{[\gamma^{i}\gamma_{5}]}(x,p_{T};T)=&\frac{M}{P^{+}} \bigg[-g_{LL}^{\perp} S_{LL}\frac{\varepsilon_{T}^{ij}p_{Tj}}{M}-g_{LT}^{\prime} \epsilon_{T}^{ij}S_{LTj}+g_{LT}^{\perp} \frac{\varepsilon_{T}^{ij}p_{Tj}S_{LT}\cdot p_{T}}{M^{2}}\nonumber\\
 &\qquad +g_{TT}^{\prime} \frac{\varepsilon_{T}^{ij}S_{TTjl}p_{T}^{l}}{M}-g_{TT}^{\perp} \frac{p_{T}\cdot S_{TT}\cdot p_{T}}{M^{2}}\frac{\varepsilon_{T}^{ij}p_{Tj}}{M} \bigg],\\
\Phi^{[i\sigma^{ij}\gamma_5]}(x,p_{T};T)=&\frac{M}{P^{+}}\bigg[-h_{LL}\epsilon_T^{ij} S_{LL}
-h_{LT}\epsilon_T^{ij}\frac{S_{LT}\cdot p_{T}}{M} 
+ h_{TT} \epsilon_T^{ij}\frac{p_{T}\cdot S_{TT}\cdot p_{T}}{M^{2}}\bigg],\\
\Phi^{[i\sigma^{-+}\gamma_5]}(x,p_{T};T)=&\frac{M}{P^{+}}\bigg[h_{LT}^{\perp}
\frac{\epsilon_T^{ij} p_{Ti}S_{LTj}}{M}
-h_{TT}^{\perp}\frac{\epsilon_T^{ij}p_{Ti}S_{TTjl}p_T^l}{M^2}\bigg].\label{e.phisigmamp}
\end{align}
In ref.~\cite{Kumano:2020ijt}, the traces are taken with $\sigma^{i+}$, $\sigma^{-+}$, and $\sigma^{ij}$, instead of $i\sigma^{i+}\gamma_5$, $i\sigma^{-+}\gamma_5$, and $i\sigma^{ij}\gamma_5$. However, the TMD PDFs obtained from both sets of traces are equivalent due to the identity 
\begin{align}
    i\sigma^{\mu\nu}\gamma_5=-\frac{1}{2}\epsilon^{\mu\nu\alpha\beta} \sigma_{\alpha\beta}.
\end{align}

In order to factor out the inhomogeneous $p_T$ dependence in eqs.~\eqref{e.phigamma}--\eqref{e.phisigmamp}, one can rewrite the terms containing multiple $p_T$ dependence.
For example, eq.~\eqref{e.sigmaip} can be written as
\begin{align}
    \Phi^{[i\sigma^{i+}\gamma_5]}(x,p_{T};T)=& h_{1LL}^{\perp}S_{LL} \frac{\epsilon_T^{ij}p_{Tj}}{M} + h_{1LT}\epsilon_T^{ij}S_{LTj}
    -h_{1LT}^\perp\frac{\epsilon_T^{ij}p_{Tjl}S_{LT}^l}{M^2}\nonumber\\
    &-h_{1TT} \frac{\epsilon_T^{ij}S_{TTjl}p_T^l}{M}
    +h_{1TT}^\perp\frac{\epsilon_T^{ij}p_{Tjlm}S_{TT}^{lm}}{M^3},\label{e.stt}
\end{align}
where $h_{1LT}$ and $h_{TT}$ are defined as
\begin{align}
    h_{1LT}&=h_{1LT}^\prime-\frac{p_T^2}{2M^2}h_{1LT}^\perp,\\
    h_{1TT}&=h_{1TT}^\prime-\frac{p_T^2}{2M^2}h_{1TT}^\perp.
\end{align}
A similar relation applies to other TMD PDFs, expressed as 
\begin{align}
    F=F^\prime -\frac{p_T^2}{2M^2}F^\perp,\label{e.Fprime}
\end{align} 
where $F$, $F^\prime$, and $F^\perp$ are independent, allowing any two to be chosen in defining the TMD PDFs in eq.~\eqref{e.TMDPDFs}.

In eq.~\eqref{e.stt}, the completely symmetric and traceless tensors $k_T^{i_1\cdots i_n}$ are~\cite{Boer:2016xqr}
\begin{align}
p_{T}^{i j}&= p_{T}^{i} p_{T}^{j}-\frac{1}{2} p_{T}^{2} g_{T}^{i j}, \label{e.kij}\\
p_{T}^{i j k}&= p_{T}^{i} p_{T}^{j} p_{T}^{k}-\frac{1}{4} p_{T}^{2}\left(g_{T}^{i j} p_{T}^{k}+g_{T}^{i k} p_{T}^{j}+g_{T}^{j k} p_{T}^{i}\right), \label{e.kijk}
\end{align}
which satisfy
$g_{T i j} k_{T}^{i j}=g_{T i j} k_{T}^{i j k}=0$.

At parton-model level, the collinear distribution functions can be obtained from the $p_T$-integrated correlation function
\begin{align}
    \Phi(x)=\int d^2\bm{p}_T \Phi(x,p_T).
\end{align}
Integrating the TMD PDFs in eq.~\eqref{e.TMDPDFs} over the transverse momentum, one can obtain three collinear PDFs for tensor-polarized components up to twist-3,
\begin{align}
    f_{1LL}(x)&=\int d^2\bm{p}_T f_{1LL}(x,p_T^2),\\ 
    e_{LL}(x)&=\int d^2\bm{p}_T e_{LL}(x,p_T^2),\\ 
    f_{LT}(x)&=\int d^2\bm{p}_T f_{LT}(x,p_T^2).
\end{align}
Consequently, the collinear correlation function can be decomposed as
\begin{align}
    \Phi(x;T)=&\frac{1}{2}\bigg\{
        f_{1LL} S_{LL} \slashed{n}_+\bigg\}
        +\frac{M}{2P^+}\bigg\{e_{LL}S_{LL} + f_{LT} \slashed{S}_{LT}\bigg\}.
\end{align}
We note that the T-odd collinear PDFs should vanish because of time-reversal invariance,
\begin{align}
    h_{1LT}(x)&=\int d^2\bm{p}_T h_{1LT}(x,p_T^2)=0,\\ 
    g_{LT}(x)&=\int d^2\bm{p}_T g_{LT}(x,p_T^2)=0,\\ 
    h_{LL}(x)&=\int d^2\bm{p}_T h_{LL}(x,p_T^2)=0.
\end{align}
We emphasize that the first function is a twist-2 collinear T-odd function: no analogous function exists in the spin-1/2 case.

The fragmentation correlation function is defined by
\begin{align}
    \Delta_{ij}(z,k_T)=\frac{1}{2z}\sum_X\int \frac{d\xi^+d^2\bm{\xi}_T}{(2\pi)^3} e^{ik\cdot \xi}
    \langle 0|\mathcal{U}^{n_+}_{(+\infty,\xi)}\psi_i(\xi)|h,X\rangle \langle h,X|\bar{\psi}_j(0)\mathcal{U}^{n_+}_{(0,+\infty)}|0\rangle \bigg|_{\xi^+=0},
\end{align}
where $k^- = P_h^-/z$, and $P_h^-$ is the dominant component of the detected hadron momentum.
The gauge link follows the definition in eq.~\eqref{e.pathint}.

Considering the production of a unpolarized hadron in the final states, we can parametrize the fragmentation correlator in terms of two twist-2 TMD FFs and four twist-3 TMD FFs,
\begin{align}
   \Delta(z,k_T)=&\frac{1}{2}\bigg\{D_1 \slashed{n}_{-}+iH_1^\perp\frac{\left[ \slashed{k}_T, \slashed{n}_{-}\right]}{2M_h}\bigg\}\nonumber\\
   &+\frac{M_h}{2P_h^-}\bigg\{E+D^\perp \frac{\slashed{k}_T}{2M_h} +iH\frac{[\slashed{n}_-,\slashed{n}_+]}{2}
   +G^\perp\gamma_5 \frac{\epsilon_T^{\rho\sigma}\gamma_\rho k_{T\sigma}}{M_h}
    \bigg\},
\end{align}
where the TMD FFs are the functions of $z$ and $k_T^2$. 
For completeness, we should add a flavor index to the TMD FFs and specify the type of the hadron in the final state.
Similarly, the collinear FFs can be obtained from the $k_T$-integrated correlation function
\begin{align}
    \Delta(z)=z^2 \int d^2\bm{k}_T \Delta(z,k_T),\label{e.deltaz}
\end{align}
where the factor $z^2$ comes from the transverse momentum $P_{hT}=-zk_T$ of the produced hadron with respect to the direction of the fragmenting quark.
The explicit parametrization of $\Delta(z)$ in terms of collinear FFs has been shown in ref.~\cite{Bacchetta:2006tn}.

The decomposition of the fragmentation correlation function $\Delta$ for tensor-polarized components can be directly obtained from $\Phi$ in eq.~\eqref{e.TMDPDFs} by the kinematic variables replacement 
\begin{align}
    n_{+} \rightarrow n_{-}, \quad \epsilon_{T} \rightarrow-\epsilon_{T}, \quad P^{+} \rightarrow P_{h}^{-}, \quad M \rightarrow M_{h}, \quad x \rightarrow 1 / z
\end{align}
and the functions notation replacement
\begin{align}
    f\rightarrow D, \quad e\rightarrow E, \quad h\rightarrow H, \quad g\rightarrow G.
\end{align}
Due to the contribution of final-state interaction between $h$ and $X$, time-reversal invariance does not imply the vanishing of T-odd collinear fragmentation functions~\cite{DeRujula:1971nnp,Hagiwara:1982cq,Jaffe:1993xb}. 
Thus, functions $H_{1LT}(z)$, $G_{LT}(z)$, and $H_{LL}(z)$ survive after integration over $k_T$.

\subsection{Quark-gluon-quark correlators}

The quark-gluon-quark correlation function is defined by~\cite{Boer:2003cm,Pijlman:2006vm}
\begin{equation}
\left(\Phi_D^\mu\right)_{ij}(x,p_T)=\int\frac{d\xi^-d^2\boldsymbol{\xi}_T}{(2\pi)^3} e^{ip\cdot\xi}\langle P|\bar{\psi}_j(0) \mathcal{U}_{(0,\xi)}^{n_-} iD^\mu(\xi) \psi_i(\xi)|P\rangle \bigg|_{\xi^+=0}, \label{e.phiD}
\end{equation}
where $iD^\mu(\xi)=i\partial^\mu+g A^\mu$ is the covariant derivative.
The plus-component of the correlator~\eqref{e.phiD} can be written as
\begin{align}
    \Phi_D^+(x,p_T)=x P^+\Phi(x,p_T)\label{e.phiDplus}
\end{align}
by using the relation $\mathcal{U}_{(0,\xi)}^{n_-}iD^+(\xi)\psi(\xi)=i\partial^+ \mathcal{U}_{(0,\xi)}^{n_-}\psi(\xi)$.
Taking the transverse component of the correlator, a further correlator $\tilde{\Phi}_A^\alpha$ can be defined as
\begin{align}
    \tilde{\Phi}_A^\alpha(x,p_T)=\Phi_D^\alpha(x,p_T)-p_T^\alpha\Phi(x,p_T), \label{e.phiA}
\end{align}
which includes only the gluon field $A^\mu$ and satisfies gauge invariance.
The correlator $\tilde{\Phi}_A^\alpha$ can be used to identify the interaction-dependent combinations in the TMD PDFs.
Since the quark-gluon-quark correlator~\eqref{e.phiD} depends on the kinematic variables of a single parton rather than two, it can not be directly decomposed as the most general form.
Correspondingly, the covariant derivative is evaluated at the same space-time point as one of the quark fields.

Under the constraint of parity conservation, the tensor-polarized part of the correlation function~\eqref{e.phiD} can be parametrized as 
\begin{align}
    \tilde{\Phi}_A^{\alpha}(x,p_T;T)& =
     \frac{x M}{2}\,
    \biggl\{ 
    \Bigl[
     \bigl(\tilde{f}_{LL}^\perp-i \tilde{g}_{LL}^{\perp}\bigr)
           S_{LL} \frac{p_{T \rho}}{M} 
    +\bigl(\tilde{f}_{LT}'-
     i \tilde{g}_{LT}'\bigr)S_{LT \rho}^{}
    -\bigl(\tilde{f}_{TT}'- i \tilde{g}_{TT}'\bigr) S_{TT \rho \sigma}^{}\frac{p_{T}^{\sigma}}{M} 
         \nonumber\\
    &-\bigl(\tilde{f}_{LT}^{\perp} -i\,\tilde{g}_{LT}^{\perp}\bigr)
         \frac{S_{LT}\cdot p_{T}}{M} \frac{p_{T \rho}^{}}{M} 
    +\bigl(\tilde{f}_{TT}^{\perp} -i\,\tilde{g}_{TT}^{\perp}\bigr)
         \frac{p_T \cdot S_{TT} \cdot p_{T}}{M^2} \frac{p_{T \rho}^{}}{M} 
     \Bigl]
    \bigl(g_T^{\alpha \rho} - i \epsilon_T^{\alpha\rho} \gamma_5\bigr)
    \nonumber\\
     & 
    -\Bigl[\bigl(\tilde{h}_{LT}^{\perp} +i\,\tilde{e}_{LT}^{\perp}\bigr)
         \frac{\epsilon_T^{\rho \sigma} S_{LT \rho} p_{T  \sigma}}{M} 
    +\bigl(\tilde{h}_{TT}^{\perp} +i\,\tilde{e}_{TT}^{\perp}\bigr)
         \frac{\epsilon_T^{\beta \rho} S_{TT \rho \sigma} p_{T \beta}p_{T}^{\sigma}}{M^2} \Bigl] \gamma_T^{\alpha}\,\gamma_5
         \nonumber\\
    & +\Bigl[
    \big(\tilde{h}_{LL}+i \tilde{e}_{LL} \big) S_{LL}
    -\bigl(\tilde{h}_{LT}- i \tilde{e}_{LT}\bigr) 
           \frac{S_{LT}\cdot p_{T}}{M} 
    -\bigl(\tilde{h}_{TT}- i \tilde{e}_{TT}\bigr) 
         \frac{p_{T}\cdot S_{TT}\cdot p_{T}}{M^2}
     \Bigr]
      i \gamma_T^{\alpha}
      \nonumber\\
    &+ \ldots \bigl(g_T^{\alpha \rho} 
                   + i \epsilon_T^{\alpha\rho} \gamma_5\bigr)
    \biggr\} \frac{\slashed{n}_+}{2} ,\label{e.PhitildeA}
    \end{align}
where $\alpha$ is a transverse indice and so forth in the following.
The functions with a tilde are interaction-dependent twist-3 distribution functions, which depend on $x$ and $p_T^2$.
The terms $\tilde{f}_{LT}^\prime$, $\tilde{g}_{LT}^\prime$, $\tilde{f}_{TT}^\prime$, and $\tilde{g}_{TT}^\prime$ have the same definition as defined in eq.~\eqref{e.Fprime}.
Note that the decomposition of the quark-gluon-quark correlator for tensor-polarized components is similar to that for a vector-polarized hadron, with the distinction arising from the parity-invariant spin tensor.
The explicit parametrization of the last term in the curly brackets is not given, as it is irrelevant to the calculation of the hadronic tensor in the SIDIS process.

The tilde functions defined in eq.~\eqref{e.PhitildeA} can be separated by the following three projection:
\begin{align}
    \frac1{2Mx}\mathrm{Tr}\left[\tilde{\Phi}_{A\alpha}\sigma^{\alpha+}\right]&=
    S_{LL}(\tilde{h}_{LL}+i\tilde{e}_{LL})
    +\frac{p_T\cdot S_{LT}}{M}(\tilde{h}_{LT}-i\tilde{e}_{LT})\nonumber\\
    &-\frac{p_T\cdot S_{TT}\cdot p_T}{M^2}(\tilde{h}_{TT}-i\tilde{e}_{TT}),\\
    \frac1{2Mx}\mathrm{Tr}\left[\tilde{\Phi}_{A\alpha} i\sigma^{\alpha+}\gamma_5\right]&=
    \frac{\epsilon_T^{\rho\sigma}S_{LT\rho}p_{T\sigma}}{M}(\tilde{h}_{LT}^\perp+i\tilde{e}_{LT}^\perp)\nonumber\\
    &+\frac{\epsilon_T^{\rho\sigma} S_{TT\rho\alpha}p_T^\alpha p_{T\sigma}}{M}(\tilde{h}_{TT}^\perp+i\tilde{e}_{TT}^\perp),\\
   \frac1{2Mx}\mathrm{Tr}\left[\tilde{\Phi}_{A\rho}(g_T^{\alpha\rho}+i\epsilon_T^{\alpha\rho}\gamma_5)\gamma^+\right]&=
   S_{LL}\frac{p_T^\alpha}{M}(\tilde{f}_{LL}^\perp-i\tilde{g}_{LL}^\perp)
   +S_{LT}^\alpha(\tilde{f}_{LT}-i \tilde{g}_{LT})\nonumber\\
   &+\frac{p_T^\alpha p_T^\rho-\frac12p_T^2g_T^{\alpha\rho}}{M^2}S_{LT\rho} (\tilde{f}_{LT}^\perp-i\tilde{g}_{LT}^\perp)
   -\frac{S_{TT}^{\alpha\rho}p_{T\rho}}{2M}(\tilde{f}_{TT}-i\tilde{g}_{TT})\nonumber\\
   &+\frac{p_T^\alpha p_T^\rho-\frac14p_T^2g_T^{\alpha\rho}}{M^2}\frac{S_{TT\rho\sigma}p_T^\sigma}{M}(\tilde{f}_{TT}^\perp-i\tilde{g}_{TT}^\perp).
\end{align}

The correlation function can be decomposed into different parts with definite twist, where the quark-quark correlator starts from twist-2,
\begin{align}
    \Phi & = \Phi_{2}+\frac{M}{P^{+}} \Phi_{3}+\left(\frac{M}{P^{+}}\right)^{2} \Phi_{4},
\end{align}
while the quark-gluon-quark correlator starts from twist-3,
\begin{align}
    \frac{1}{M} \tilde{\Phi}_{A}^{\alpha} & = \tilde{\Phi}_{A, 3}^{\alpha}+\frac{M}{P^{+}} \tilde{\Phi}_{A, 4}^{\alpha}+\left(\frac{M}{P^{+}}\right)^{2} \tilde{\Phi}_{A, 5}^{\alpha}.
\end{align}
Here the number in the subscript indicates the specific twist of TMD PDFs.
For simplicity, we suppress the dependence of $x$ and $p_T$.
The QCD equation of motion can connect the correlation functions of different twist.
In light-cone coordinates, it is written as
\begin{align}
    [i \slashed{D}(\xi)-m] \psi(\xi) 
    = \left[\gamma^{+} i D^{-}(\xi)+\gamma^{-} i D^{+}(\xi)+\gamma_{T}^{\alpha} i D_{\alpha}(\xi)-m\right] \psi(\xi) = 0,\label{e.EoM}
\end{align}
where $m$ is the mass of the quark.
With the projection operators $\mathcal{P}_+=\frac{1}{2}\gamma^- \gamma^+$ and $\mathcal{P}_-=\frac{1}{2} \gamma^+ \gamma^-$~\cite{Jaffe:1996zw}, we have the identities $\mathcal{P}_+ \Phi_4=\mathcal{P}_-\Phi_2=0$ and $\mathcal{P}_+ \tilde{\Phi}_{A,5}^\alpha=\mathcal{P}_-\tilde{\Phi}_{A,3}^\alpha=0$. Using these relations,
we can project out the good component of eq.~\eqref{e.EoM} and express the corresponding correlators as
\begin{align}
    \mathcal{P}_{+}\left[x M \gamma^{-} \Phi_{3}+M \gamma_{T \rho}\right. & \left.\tilde{\Phi}_{A, 3}^{\rho}+\slashed{p}_{T} \Phi_{2}-m \Phi_{2}\right] \nonumber\\
    & +\mathcal{P}_{+}\left[x M \gamma^{-} \Phi_{4}+M \gamma_{T \rho} \tilde{\Phi}_{A, 4}^{\rho}+\slashed{p}_{T} \Phi_{3}-m \Phi_{3}\right] \frac{M}{P^{+}} = 0,\label{e.PplusEoM}
\end{align}
where the term with $D^-$ in eq.~\eqref{e.EoM} does not contribute and the terms $D^+$ and $D_{\alpha}$ are replaced by eqs.~\eqref{e.phiDplus} and~\eqref{e.phiA}, respectively.
Projecting this relation on specific Dirac structures and taking the trace gives
\begin{align}
   \mathrm{Tr} \Gamma^+\left[x \gamma^- \Phi_3+\gamma_{T\rho} \tilde{\Phi}_{A,3}^\rho+\frac{\slashed{p}_T}{M} \Phi_2-\frac{m}{M} \Phi_2\right]=0, \label{e.difftwist}
\end{align}
where $\Gamma^+$ is one of the matrices $\Gamma^+=\{\gamma^+,\gamma^+\gamma_5,i\sigma^{\alpha+}\gamma_5\}$.
We find that the terms with factor $M/P^{+}$ in eq.~\eqref{e.PplusEoM} vanish because the trace of Dirac matrices cannot produce a term that transforms as $P^+$ under boosts along the light-cone direction.

Substituting the expression of the quark-quark correlator~\eqref{e.TMDPDFs} containing $\Phi_2$ and $\Phi_3$ and the quark-gluon-quark correlator $\tilde{\Phi}_A^\alpha$~\eqref{e.PhitildeA} into eq.~\eqref{e.difftwist}, one can establish the relations between the twist-2 and twist-3 functions.
For a tensor-polarized spin-1 hadron, we derive 12 relations between T-even functions,
\begin{align}
    &x e_{LL}=x \tilde{e}_{LL}+\frac{m}{M}f_{1LL},\label{e.xeLL}\\
    &x f^{\perp}_{LL} =x  \tilde{f}^{\perp}_{LL}+ f_{1LL},\\
    &x e_{LT}^\perp=x \tilde{e}_{LT}+\frac{m}{M}f_{1LT},\label{e.eLT}\\
    &x e_{TT}^\perp=x \tilde{e}_{TT}+\frac{m}{M}f_{1TT},\label{e.eTT}\\
    &x e_{LT}=x \tilde{e}_{LT}^\perp,\\
    & x e_{TT}=x \tilde{e}_{TT}^\perp,\label{e.eTTperp}\\
    &x f_{LT}^\prime=x\tilde{f}_{LT}^\prime,\\
    & x f_{LT}=x\tilde{f}_{LT}-\frac{p_T^2}{2M^2}f_{1LT},\\
    & x f_{LT}^\perp=x\tilde{f}_{LT}^\perp+f_{1LT},\\
    & x f_{TT}^\prime=x\tilde{f}_{TT}^\prime,\\
    & x f_{TT}=x\tilde{f}_{TT}-\frac{p_T^2}{2M^2}f_{1TT},\\
    &x f_{TT}^\perp=x\tilde{f}_{TT}^\perp+f_{1TT},
\end{align}
and 12 relations between T-odd functions,
\begin{align}
    &x g^{\perp}_{LL} =  x  \tilde{g}^{\perp}_{LL}+\frac{m}{M} h_{1LL}^{\perp},\\
    &  x g_{LT}^\prime=x \tilde{g}_{LT}^\prime + \frac{p_T^2}{M^2}g_{1LT} +\frac{m}{M}h_{1LT} + \frac{m}{M}\frac{p_T^2}{2M^2}h_{1LT}^\perp,\\
    &x g_{LT}^\perp=x \tilde{g}_{LT}^\perp + g_{1LT} + \frac{m}{M}h_{1LT}^\perp,\\
    &  x g_{LT}=x \tilde{g}_{LT}-\frac{p_T^2}{2M^2}g_{1LT}+\frac{m}{M}h_{1LT},\\
    & x g_{TT}^\prime=x \tilde{g}_{TT}^\prime - \frac{p_T^2}{M^2}g_{1TT} +\frac{m}{M}h_{1TT}+ \frac{m}{M}\frac{p_T^2}{2M^2}h_{1TT}^\perp,\\
    &x g_{TT}^\perp=x \tilde{g}_{TT}^\perp - g_{1TT} + \frac{m}{M}h_{1TT}^\perp,\\
    & x g_{TT}=x \tilde{g}_{TT}- \frac{p_T^2}{2M^2}g_{1TT} +\frac{m}{M}h_{1TT},\\
    &x h_{LL}  =x  \tilde{h}_{LL} + \frac{p_T^2}{M^2}\, h_{1LL}^{\perp}.\label{e.xhLL}\\
    & x h_{LT}=x \tilde{h}_{LT}+h_{1LT}-\frac{p_T^2}{2M^2}h_{1LT}^\perp,\\
    & x h_{LT}^\perp=x \tilde{h}_{LT}^\perp+h_{1LT}+\frac{p_T^2}{2M^2}h_{1LT}^\perp+\frac{m}{M}g_{1LT},\\
    & x h_{TT}=x\tilde{h}_{TT}+h_{1TT}-\frac{p_T^2}{2M^2}h_{1TT}^\perp,\\
    & x h_{TT}^\perp=x \tilde{h}_{TT}^\perp+h_{1TT}+\frac{p_T^2}{2M^2}h_{1TT}^\perp-\frac{m}{M}g_{1TT},\label{e.xhTTp}
\end{align}
where we kept the sign of the functions with a tilde in eq.~\eqref{e.PhitildeA} consistent with those defined in the quark-quark correlator~\eqref{e.TMDPDFs}. 
The functions with a tilde in eqs.~\eqref{e.xeLL}-\eqref{e.xhTTp} vanish in the so-called Wandzura--Wilczek approximation, which is equivalent to neglecting the quark-gluon-quark correlator (see, e.g., refs.~\cite{Wandzura:1977qf,Accardi:2009au,Bastami:2018xqd}).
We can find that the $S_{LL}$-dependent relations share the same forms as the unpolarized ones in ref.~\cite{Bacchetta:2006tn}.
Each of the TMD PDFs defined in eq.~\eqref{e.TMDPDFs} have the corresponding relations derived from the equation of motion, except for four additional ones specified by the identity~\eqref{e.Fprime}.

The analysis above can be applied to the fragmentation part, which has been discussed in ref.~\cite{Bacchetta:2006tn}. We list the results below for later use.
Similar to $\Phi_D$, the quark-gluon-quark fragmentation correlation function is defined as
\begin{align}
&\left(\Delta_D^\mu\right)_{ij}(z,k_T)=\nonumber\\
&\frac1{2z}\sum_X\int\frac{d\xi^+}{(2\pi)^3}\left.e^{ik\cdot\xi}\left\langle0\right|\mathcal{U}_{(+\infty,\xi)}^{n_+}iD^\mu(\xi)\left.\psi_i(\xi)|h,X\right\rangle\langle h,X|\bar{\psi}_j\left(0\right)\mathcal{U}_{(0,+\infty)}^{n_+}|0\rangle\right|_{\xi^-=0}.
\end{align}
Restricting to the transverse component, we have
\begin{align}
    \tilde{\Delta}_A^\alpha(z,k_T) = \Delta_D^\alpha(z,k_T) -k_T^\alpha\Delta(z,k_T),   
\end{align}
which can be decomposed as~\cite{Bacchetta:2006tn}
\begin{align}
\tilde{\Delta}_A^\alpha(z,k_T)=&
\frac{M_h}{2z}\bigg\{\big(\tilde{D}^\perp-i \tilde{G}^\perp\big)\frac{k_{T\rho}}{M_h}\left(g_T^{\alpha\rho}+i\epsilon_T^{\alpha\rho}\gamma_5\right)\nonumber\\
&\qquad+\big(\tilde{H}+i \tilde{E}\big)i\gamma_T^\alpha+\ldots\left(g_T^{\alpha\rho}-i\epsilon_T^{\alpha\rho}\gamma_5\right)\bigg\}\frac{\slashed{n}_-}{2}.
\end{align}
According to the equation of motion for the quark field, the relations between the functions from the quark-quark correlator and the functions from the quark-gluon-quark correlator can be established,
\begin{align}
\frac{E}{z}& =\frac{\tilde{E}}{z}+\frac{m}{M_h} D_1, \\
\frac{D^\perp}{z}& =\frac{\tilde{D}^\perp}z+D_1, \\
\frac{G^\perp}z& =\frac{\tilde{G}^\perp}z+\frac m{M_h} H_1^\perp, \\
\frac{H}z& =\frac{\tilde{H}}z+\frac{k_T^2}{M_h^2} H_1^\perp.
\end{align}

\subsection{Results for structure functions}
\label{s.strucfuncresults}

Substituting the correlators in the hadronic tensor~\eqref{e.subwmunu} with the parametrization of the different correlators and using the relations from the equation of motion, we can derive the complete expression of the hadronic tensor in the parton model, as presented in Appendix~\ref{a.hadrontensor}.
After contracting the hadronic tensor obtained above and the leptonic tensor, one can express the structure functions appearing in eq.~\eqref{e.dsigma} in terms of the convolution of TMD PDFs and FFs up to twist-3.
To obtain more concise expressions, we introduce a transverse momentum convolution notation
\begin{align}
	\mathcal{C}[w f D]=x \sum_a e_a^2 \int d^2 \boldsymbol{k}_T d^2 \boldsymbol{p}_T \delta^{(2)}\left(\boldsymbol{k}_T-\boldsymbol{p}_T-\boldsymbol{P}_{h \perp} / z\right) w\left(\boldsymbol{k}_T, \boldsymbol{p}_T\right) f^a\left(x, k_T^2\right) D^a\left(z, p_T^2\right),
\end{align}
where $f^a$ and $D^a$ indicate a TMD PDF for the tensor-polarized target and a TMD FF for the detected hadron, respectively. 
The $w\left(\boldsymbol{k}_T, \boldsymbol{p}_T\right)$ is one of the dimensionless scalar functions of $\boldsymbol{k}_T$ and $\boldsymbol{p}_T$.
The results for the structure functions are given by       
\begin{align}
    &F_{U(LL),T}=\mathcal{C}[f_{1LL}D_1],\label{e.FULLT}\\
    &F_{U(LL),L}=0,\label{e.FULLL}\\
    &F_{U(LL)}^{\cos\phi_{h}}
    =\frac{2M}{Q}\mathcal{C}\bigg[
    -\frac{\hat{\bm{h}}\cdot\bm{p}_{T}}{M}\bigg(xf_{LL}^{\perp}D_{1}+\frac{M_{h}}{M}h_{1LL}^{\perp}\frac{\tilde{H}}{z}\bigg)
    -\frac{\hat{\bm{h}}\cdot\bm{k}_{T}}{M_{h}}\bigg(xh_{LL}H_{1}^{\perp}+\frac{M_{h}}{M}f_{1LL}\frac{\tilde{D}^{\perp}}{z}\bigg)\bigg],\label{e.FULLcosphi}\\
    &F_{U(LL)}^{\cos{2\phi_h}}=\mathcal{C}\bigg[-\frac{2(\hat{\bm{h}}\cdot\bm{k}_T)(\hat{\bm{h}}\cdot\bm{p}_T)-\bm{k}_T\cdot\bm{p}_T}{MM_h}h_{1LL}^\perp H_1^\perp \bigg],\\
    &F_{L(LL)}^{\sin\phi_{h}}
    =\frac{2M}{Q}\mathcal{C}\bigg[-\frac{\hat{\bm{h}}\cdot\bm{k}_{T}}{M_{h}}\bigg(xe_{LL}H_{1}^{\perp}+\frac{M_{h}}{M}f_{1LL}\frac{\tilde{G}^{\perp}}{z}\bigg)+\frac{\hat{\bm{h}}\cdot\bm{p}_{T}}{M}\bigg(xg_{LL}^{\perp}D_{1}+\frac{M_{h}}{M}h_{1LL}^{\perp}\frac{\tilde{E}}{z}\bigg)\bigg],\label{e.FLLLsinphi}\\
    &F_{U(LT),T}^{\cos{(\phi_h}-\phi_{LT})}=\mathcal{C}\bigg[\frac{\hat{\bm{h}}\cdot\bm{p}_T}{M} f_{1LT}D_1 \bigg],\\
    &F_{U(LT),L}^{\cos{(\phi_h}-\phi_{LT})}=0,\label{e.FULTL}\\
    &F_{U(LT)}^{\cos\phi_{LT}}
    =\frac{2M}{Q}\mathcal{C}\Bigg\{-\bigg[\bigg(xf_{LT}D_{1}+\frac{M_{h}}{M}h_{1LT}\frac{\tilde{H}}{z}\bigg)\bigg]\nonumber\\
    &\qquad\qquad+\frac{\bm{k}_{T}\cdot\bm{p}_{T}}{2MM_{h}}\bigg[\bigg(xh_{LT}H_{1}^{\perp}+\frac{M_{h}}{M}g_{1LT}\frac{\tilde{G}^{\perp}}{z}\bigg)+\bigg(xh_{LT}^{\perp}H_{1}^{\perp}-\frac{M_{h}}{M}f_{1LT}\frac{\tilde{D}^{\perp}}{z}\bigg)\bigg]\Bigg\},\\
    &F_{U(LT)}^{\cos(2\phi_{h}-\phi_{LT})}
    =\frac{2M}{Q}\mathcal{C}\Bigg\{\frac{\bm{p}_{T}^{2}-2(\hat{\bm{h}}\cdot\bm{p}_{T})^{2}}{2M^{2}}\bigg(xf_{LT}^{\perp}D_{1}+\frac{M_{h}}{M}h_{1LT}^{\perp}\frac{\tilde{H}}{z}\bigg)\nonumber\\
    &\qquad\qquad\qquad-\frac{2(\hat{\bm{h}}\cdot\bm{k}_{T})(\hat{\bm{h}}\cdot\bm{p}_{T})-\bm{k}_{T}\cdot\bm{p}_{T}}{2MM_{h}}\bigg[\bigg(xh_{LT}^{\perp}H_{1}^{\perp}+\frac{M_{h}}{M}f_{1LT}\frac{\tilde{D}^{\perp}}{z}\bigg)\nonumber\\
    &\qquad\qquad\qquad-\bigg(xh_{LT}H_{1}^{\perp}-\frac{M_{h}}{M}g_{1LT}\frac{\tilde{G}^{\perp}}{z}\bigg)\bigg]\Bigg\},\label{e.(4.90)}\\
    &F_{U(LT)}^{\cos{(\phi_h+\phi_{LT})}}=\mathcal{C}\bigg[-\frac{\hat{\bm{h}}\cdot\bm{k}_T}{M_h}h_{1LT}H_1^\perp \bigg],\\
    &F_{U(LT)}^{\cos{(3\phi_h}-\phi_{LT})}=\mathcal{C}\bigg[-\frac{4(\hat{\bm{h}}\cdot\bm{k}_T)(\hat{\bm{h}}\cdot\bm{p}_T)^2-(\hat{\bm{h}}\cdot\bm{k}_T)\bm{p}_T^2-2(\hat{\bm{h}}\cdot\bm{p}_T)(\bm{k}_T\cdot\bm{p}_T) }{2M^2M_h}h_{1LT}^\perp H_1^\perp\bigg],\label{e.(4.92)}\\
    &F_{L(LT)}^{\sin\phi_{LT}}
    =\frac{2M}{Q}\mathcal{C}\Bigg\{\bigg(xg_{LT}D_{1}+\frac{M_{h}}{M}h_{1LT}\frac{\tilde{E}}{z}\bigg)\nonumber\\
    &\qquad\qquad+\frac{\bm{k}_{T}\cdot\bm{p}_{T}}{2MM_{h}}\bigg[\bigg(xe_{LT}H_{1}^{\perp}-\frac{M_{h}}{M}g_{1LT}\frac{\tilde{D}^{\perp}}{z}\bigg)-\bigg(xe_{LT}^{\perp}H_{1}^{\perp}+\frac{M_{h}}{M}f_{1LT}\frac{\tilde{G}^{\perp}}{z}\bigg)\bigg]\Bigg\},\\
    &F_{L(LT)}^{\sin(2\phi_{h}-\phi_{LT})}
    =\frac{2M}{Q}\mathcal{C}\Bigg\{-\frac{2(\hat{\bm{h}}\cdot\bm{k}_{T})(\hat{\bm{h}}\cdot\bm{p}_{T})-\bm{k}_{T}\cdot\bm{p}_{T}}{2MM_{h}}\bigg[\bigg(xe_{LT}H_{1}^{\perp}+\frac{M_{h}}{M}f_{1LT}\frac{\tilde{G}^{\perp}}{z}\bigg)\nonumber\\
    &\qquad\qquad\qquad+\bigg(xe_{LT}^{\perp}H_{1}^{\perp}-\frac{M_{h}}{M}g_{1LT}\frac{\tilde{D}^{\perp}}{z}\bigg)\bigg]\Bigg\},\\
    &F_{L(LT)}^{\sin{(\phi_h-\phi_{LT}})}=\mathcal{C}\bigg[-\frac{\hat{\bm{h}}\cdot\bm{p}_T}{M} g_{1LT}D_1 \bigg],\\
    &F_{U(TT),T}^{\cos{(2\phi_h-2\phi_{LT})}}=\mathcal{C}\bigg[-\frac{2(\hat{\bm{h}}\cdot\bm{p}_T)^2-\bm{p}_T^2}{M^2}f_{1TT}D_1\bigg],\\
    &F_{U(TT),L}^{\cos{(2\phi_h-2\phi_{LT})}}=0,\\
    &F_{U(TT)}^{\cos(\phi_{h}-2\phi_{TT})}
    =\frac{2M}{Q}\mathcal{C}\Bigg\{\frac{\hat{\bm{h}}\cdot\bm{p}_{T}}{M}\big(xf_{TT}D_{1}-\frac{M_{h}}{M}h_{1TT}\frac{\tilde{H}}{z}\big)\nonumber\\
    &\qquad\qquad\qquad+\frac{(\hat{\bm{h}}\cdot\bm{k}_{T})\bm{p}_{T}^{2}-2(\hat{\bm{h}}\cdot\bm{p}_{T})\big(\bm{k}_{T}\cdot\bm{p}_{T}\big)}{2M^{2}M_{h}}\bigg[\bigg(xh_{TT}H_{1}^{\perp}-\frac{M_{h}}{M}g_{1TT}\frac{\tilde{G}^{\perp}}{z}\bigg)\nonumber\\
    &\qquad\qquad\qquad+\bigg(xh_{TT}^{\perp}H_{1}^{\perp}-\frac{M_{h}}{M}f_{1TT}\frac{\tilde{D}^{\perp}}{z}\bigg)\bigg]\Bigg\},\\
    &F_{U(TT)}^{\cos(3\phi_{h}-2\phi_{TT})}
    =\frac{2M}{Q}\mathcal{C}\Bigg\{\frac{3(\hat{\bm{h}}\cdot\bm{p}_{T})\big(2(\hat{\bm{h}}\cdot\bm{p}_{T})^{2}-\bm{p}_{T}^{2}\big)}{2M^{3}}\bigg(xf_{TT}^{\perp}D_{1}+\frac{M_{h}}{M}h_{1TT}^{\perp}\frac{\tilde{H}^{\perp}}{z}\bigg)\nonumber\\
    &\qquad\qquad\qquad+\frac{4(\bm{h}\cdot\bm{k}_{T})(\bm{h}\cdot\bm{p}_{T})^{2}-2(\bm{h}\cdot\bm{p}_{T})(\bm{k}_{T}\cdot\bm{p}_{T})-(\bm{h}\cdot\bm{k}_{T})\bm{p}_{T}^{2}}{2M^{2}M_{h}}\nonumber\\
    &\qquad\qquad\qquad \times \bigg[\bigg(xh_{TT}^{\perp}H_{1}^{\perp}+\frac{M_{h}}{M}f_{1TT}\frac{\tilde{D}^{\perp}}{z}\bigg)
    -\bigg(xh_{TT}H_{1}^{\perp}+\frac{M_{h}}{M}g_{1TT}\frac{\tilde{G}^{\perp}}{z}\bigg)\bigg]\Bigg\},\\
    &F_{U(TT)}^{\cos{(2\phi_{TT})}}=\mathcal{C}\bigg[\frac{\bm{k}_T\cdot\bm{p}_T}{MM_h} h_{1TT}H_1^\perp \bigg],\\
    &F_{U(TT)}^{\cos{(4\phi_h-2\phi_{TT})}}=\mathcal{C}\bigg[
        -\bigg(\frac{4(\hat{\bm{h}}\cdot\bm{k}_T)(\hat{\bm{h}}\cdot\bm{p}_T)\bm{p}_T^2-8(\hat{\bm{h}}\cdot\bm{k}_T)(\hat{\bm{h}}\cdot\bm{p}_T)^3}{2M^3M_h}\nonumber\\
    &\qquad\qquad\qquad+\frac{4(\bm{k}_T\cdot\bm{p}_T)(\hat{\bm{h}}\cdot\bm{p}_T)^2-(\bm{k}_T\cdot\bm{p}_T)\bm{p}_T^2}{2M^3 M_h}\bigg) h_{1TT}^\perp H_1^\perp \bigg],\\
    &F_{L(TT)}^{\sin(\phi_{h}-2\phi_{TT})}
    =\frac{2M}{Q}\mathcal{C}\Bigg\{\frac{\hat{\bm{h}}\cdot\bm{p}_{T}}{M}\bigg(xg_{TT}D_{1}+\frac{M_{h}}{M}h_{1TT}\frac{\tilde{E}}{z}\bigg)\nonumber\\
    &\qquad\qquad\qquad-\frac{(\hat{\bm{h}}\cdot\bm{k}_{T})\bm{p}_{T}^{2}-2(\hat{\bm{h}}\cdot\bm{p}_{T})(\bm{k}_{T}\cdot\bm{p}_{T})}{2M^{2}M_{h}}\bigg[\bigg(xe_{TT}H_{1}^{\perp}-\frac{M_{h}}{M}f_{1TT}\frac{\tilde{G}^{\perp}}{z}\bigg)\nonumber\\
    &\qquad\qquad\qquad-\bigg(xe_{TT}^{\perp}H_{1}^{\perp}-\frac{M_{h}}{M}g_{1TT}\frac{\tilde{D}^{\perp}}{z}\bigg)\bigg]\Bigg\},\\
    &F_{L(TT)}^{\sin(3\phi_{h}-2\phi_{TT})}
    =\frac{2M}{Q}\mathcal{C}\Bigg\{-\frac{3(\hat{\bm{h}}\cdot\bm{p}_{T})\big(2(\hat{\bm{h}}\cdot\bm{p}_{T})^{2}-\bm{p}_{T}^{2}\big)}{2M^{3}}\bigg(xg_{TT}^{\perp}D_{1}+\frac{M_h}{M}h_{1TT}^{\perp}\frac{\tilde{E}}{z}\bigg)\nonumber\\
    &\qquad\qquad\qquad+\frac{4(\bm{h}\cdot\bm{k}_{T})(\bm{h}\cdot\bm{p}_{T})^{2}-2(\bm{h}\cdot\bm{p}_{T})(\bm{k}_{T}\cdot\bm{p}_{T})-(\bm{h}\cdot\bm{k}_{T})\bm{p}_{T}^{2}}{2M^{2}M_{h}}\nonumber\\
    &\qquad\qquad\qquad\times
    \bigg[\bigg(xe_{TT}H_{1}^{\perp}+\frac{M_h}{M}f_{1TT}\frac{\tilde{G}^{\perp}}{z}\bigg)
    +\bigg(xe_{TT}^{\perp}H_{1}^{\perp}+\frac{M_h}{M}g_{1TT}\frac{\tilde{D}^{\perp}}{z}\bigg)\bigg]\Bigg\},\\
    &F_{L(TT)}^{\sin{(2\phi_h-2\phi_{TT}})}=\mathcal{C}\bigg[-\frac{2(\hat{\bm{h}}\cdot\bm{p}_T)^2-\bm{p}_T^2}{M^2}g_{1TT}D_1 \bigg],\label{e.FLTT}
\end{align}
where the unit vector $\hat{\bm{h}}=\bm{P}_{h\perp}/|\bm{P}_{h\perp}|$ gives the direction of the transverse momentum.
The $S_{LL}$-dependent structure functions in eqs.~\eqref{e.FULLT}-\eqref{e.FLLLsinphi} share the same expressions as those for unpolarized states given by eqs.~(4.1)-(4.6) in ref.~\cite{Bacchetta:2006tn}.

Up to twist-3, only two structure functions are zero, given by eqs.~\eqref{e.FULLL} and \eqref{e.FULTL}.
However, if we limit ourselves to the twist-2 level, 12 out of the 23 structure functions vanish. 
We should also note that structure functions involving twist-2 terms exhibit dependence on an even number of azimuthal angles ($\phi_h$, $\phi_{LT}$, $\phi_{TT}$), whereas twist-3 structure functions depend on an odd number of azimuthal angles.
For example, the $F_{U(LT)}^{\cos (2\phi_h-\phi_{LT})}$ in eq.~\eqref{e.(4.90)} depends on $2\phi_h$ and $\phi_{LT}$, which sum to three (odd) azimuthal angles, and then it receives the leading contribution at twist-3. The $F_{U(LT)}^{\cos (3\phi_h-\phi_{LT})}$ in eq.~\eqref{e.(4.92)} depends on $3\phi_h$ and $\phi_{LT}$, which sum to four (even) azimuthal angles, and then it receives the leading contribution at twist-2.
These nonvanishing structure functions can be utilized to study the tensor-polarized TMD PDFs for a spin-1 particle.

For twist-2 structure functions, the inclusion of higher-order QCD corrections leads to two main changes to the results presented here: the TMD PDFs and FFs will acquire a dependence on two scales, usually denoted by $\mu$ and $\sqrt{\zeta}$, the convolution in the structure functions must be multiplied by a hard factor that depends on $Q^2$ and $\mu$. TMD factorization guarantees that the complete structure functions depend only on $Q^2$ and leads to specific evolution equations for the involved TMDs~\cite{Collins:2011zzd}. The effect of higher-order corrections on twist-3 structure functions has been studied recently and leads to similar changes~\cite{Bacchetta:2019qkv, Ebert:2021jhy, Rodini:2023plb}.

Integrating eqs.~\eqref{e.FULLT}-\eqref{e.FLTT} over the transverse momentum of the produced hadron, $\bm{P}_{h\perp}$, one can express the integrated structure functions in eq.~\eqref{e.dsigmadz} as
\begin{align}
    F_{U(LL),T}(x_d,z)&=x \sum_a e_a^2 f_{1LL}^a(x) D_1^a(z),\\
    F_{U(LL),L}(x_d,z)&=0,\\
    F_{U(LT)}^{\cos\phi_{LT}}(x_d,z)&=-x\sum_a e_a^2 \frac{2M}{Q} f_{LT}^a(x) D_1^a(z),
    \label{e.cosphiLT}\\
    F_{U(TT)}^{\cos(2\phi_{TT})}(x_d,z)&=0,
\end{align}
where the T-odd functions vanish after the integration over the transverse momentum.\footnote{If collinear T-odd functions were included in the analysis, eq.~\eqref{e.cosphiLT} would become
$F_{U(LT)}^{\cos\phi_{LT}}=-\sum_a e_a^2 \frac{2M}{Q} \Bigl(x f_{LT}^a D_1^a+\frac{M_h}{M}h_{1LT}^a \frac{\tilde{H}^a}{z}\Bigr)$
and a new structure function $F_{U(LT)}^{\sin\phi_{LT}}=\sum_a e_a^2 \frac{2M}{Q} \Bigl(x g_{LT}^a D_1^a+\frac{M_h}{M}h_{1LT}^a \frac{\tilde{E}^a}{z}\Bigr)$ would appear.}
\begin{align}
    F_{U(LL),T}(x_d)&=x \sum_a e_a^2 f_{1LL}^a(x),\\
    F_{U(LL),L}(x_d)&=0,\\
    F_{U(LT)}^{\cos\phi_{LT}}(x_d)&=-x\sum_a e_a^2 \frac{2M}{Q} f_{LT}^a(x),\\
    F_{U(TT)}^{\cos(2\phi_{TT})}(x_d)&=0,
\end{align} 
where we use the sum rule $\displaystyle{\sum_h} \int dz\,z\, D_1^a(z) =1$.
We find that only two structure functions contribute to the inclusive DIS cross section up to twist-3 and can be used to investigate the tensor-polarized dependent collinear PDFs for a spin-1 particle.
According to eqs.~\eqref{e.brelation1}-\eqref{e.brelation4}, we can express $b_{1-4}$ as (cf. ref.~\cite{Cosyn:2024drt})
\begin{align}
    b_1&=-\frac{1}{2(\gamma^2+1)}\Big(\frac{3}{2}F_{U(LL),T}+ F_{U(TT)}^{\cos(2\phi_{TT})} \Big),\label{e.b1}\\
    b_2&=-\frac{x_d}{(\gamma^2+1)^2}\Big[\frac{3}{2}\big(F_{U(LL),L}+F_{U(LL),T}\big)
    +4\gamma F_{U(LT)}^{\cos\phi_{LT}}+(2\gamma^2+1) F_{U(TT)}^{\cos(2\phi_{TT})} \Big],\\
    \gamma^2 b_3&=\frac{x_d}{3(\gamma^2+1)^2}\Big[-\frac{3}{2}\gamma^2\big( F_{U(LL),L}+ F_{U(LL),T} \big)-4\gamma^3 F_{U(LT)}^{\cos\phi_{LT}}\nonumber\\
    &\qquad+(4\gamma^4+11\gamma^2+6)F_{U(TT)}^{\cos(2\phi_{TT})} \Big],\\
    \gamma^2 b_4&=-\frac{x_d}{3(\gamma^2+1)^2}\Big[\frac{3}{2}
    \gamma^2 \big(F_{U(LL),L}+ F_{U(LL),T}\big)-4\gamma(2\gamma^2+3)F_{U(LT)}^{\cos\phi_{LT}}\nonumber\\
    &\qquad+(2\gamma^4+13\gamma^2+12)F_{U(TT)}^{\cos(2\phi_{TT})}  \Big].\label{e.b4}
\end{align}

Compared to the SIDIS process off a spin-1/2 target, many new structure functions are introduced for tensor polarized spin-1 target. While all these structure functions are in principle measurable in experiments via corresponding spin asymmetries and the separation of various azimuthal modulations, we would like to discuss and emphasize some particular terms considering experimental feasibility and physical insights. 

In general, leading-twist observables are experimentally favorable, because higher-twist contributions are formally suppressed in powers of $1/Q$ and are therefore more challenging. 
Measurements of the structure functions containing $D_1$, such as $F_{U(LL),T}$, $F_{U(LT),T}^{\cos(\phi_h-\phi_{LT})}$, $F_{L(LT)}^{\sin(\phi_h-\phi_{LT})}$, $F_{U(TT),T}^{\cos(2\phi_h-2\phi_{LT})}$, and $F_{L(TT)}^{\sin(2\phi_h-2\phi_{TT})}$ should provide the clearest access to tensor-polarized TMDs, as they are convoluted with the relatively well-constrained unpolarized fragmentation function $D_1$. The kinematic coverage available at JLab and at the future EIC makes these leading-twist channels especially promising for early measurements.

The $S_{LL}$ sector is expected to be the most accessible experimentally. An $S_{LL}$-polarized deuteron target has already been realized in the HERMES measurement~\cite{HERMES:2005pon}, and a new experiment at JLab has been proposed to measure the deuteron tensor structure function with an $S_{LL}$-polarized target~\cite{Jlabproposal}. The tensor-polarized PDFs and TMDs can probe new feature of the deuteron structure beyond a simple bound system of a proton and a neutron. In particular, the large difference between the HERMES data and the standard deuteron model calculations indicates such a possibility. The structure function $b_1$ in inclusive scattering remains a primary aim and is a key measurement of the proposed experiment at JLab~\cite{Jlabproposal}. Its counterpart in SIDIS, $F_{U(LL),T}$, is also of particular interest, and dedicated analyses for extracting $f_{1LL}$ have been proposed~\cite{Poudel:2025nof,Poudel:2025tac}. Compared to inclusive DIS, the transverse-momentum $P_{h\perp}$ dependence in SIDIS can discriminate among competing models for $b_1$, and the convolution with TMD fragmentation functions offers sensitivity to the flavor separation.
Besides, the azimuthal dependent term $F_{U(LL)}^{\cos 2\phi_h}$, which involves the only other leading-twist $S_{LL}$-dependent TMD PDF $h_{1LL}$, also deserves attention.

For the $S_{LT}$ sector, in addition to the leading-twist terms, the twist‑3 modulation term $F_{U(LT)}^{\cos\phi_{LT}}$ is also noteworthy, because it survives upon integrating over $P_{h\perp}$ and even contributes in inclusive DIS. The $\sin\phi_{LT}$ modulation term is naively T‑odd and should vanish after the integration over $P_{h\perp}$, providing a test on the role of intrinsic transverse momentum and gauge‑link–induced phases in generating T‑odd effects~\cite{Bacchetta:2002xd}.

For a realistic demonstration of experimental feasibility, which relies on the expected magnitude of corresponding asymmetries, model calculations are desired.
The traditional convolution model is a commonly used approach for describing the deuteron structure~\cite{Hoodbhoy:1988am,Frankfurt:1983qs, Cosyn:2017fbo, Khan:1991qk}. Although its results differ from those of the HERMES data, one may improve it and perform numerical estimates for the tensor-polarized structure functions, such as by extending it to the partonic level.
Additionally, the calculation of the tensor polarized TMD PDFs for the deuteron is underway by utilizing light-front wave functions~\cite{Sargsian:2022rmq,Kaur:2025css,Mondal:2025fdl}.
These model estimates will provide some quantitative guidance for future experimental measurements of the deuteron structures.

\section{Summary}
\label{s.summary}
This work examines the semi-inclusive deep inelastic scattering off a tensor-polarized spin-1 target. Considering a longitudinal polarized lepton beam and the production of an unpolarized hadron in this process, we derive the general form of the cross section in terms of structure functions.
Through a detailed kinematic analysis, we express the hadronic tensor as a combination Lorentz tensors and scalar coefficients.
As a result, the tensor-polarized sector of the SIDIS cross section can be decomposed using 23 structure functions, which depend the target's spin configurations and azimuthal angles. By integrating the SIDIS cross section, we also obtain the structure functions for inclusive DIS.

In the kinematic region where $\bm{q}_T^2\ll Q^2$, we 
compute the structure functions in the parton model up to the subleading order in $1/Q$ (twist-3).
We first give the expression of the hadronic tensor in terms of the transverse-momentum-dependent quark-quark correlator and quark-gluon-quark correlator, which can be parametrized in terms of the TMD PDFs and TMD FFs up to twist-3.
With the constraint of the equation of motion, the relations between the quark-quark correlator and quark-gluon-quark correlator are established.
Using these relations and the parametrization of the correlators, we express the structure functions appearing in the cross section as a convolution of the TMD PDFs and TMD FFs.
At the leading and first subleading twist accuracy, we find that all but two structure functions have nontrivial expressions in SIDIS, whereas only two structure functions contribute to the cross section in inclusive DIS.

These nonzero structure functions serve as critical tools for investigating the tensor-polarized properties of spin-1 particles. Upcoming experiments, such as those at JLab using tensor-polarized deuteron targets, will facilitate measurements, enhancing our understanding of nucleon spin dynamics and potential novel nuclear phenomena.

\acknowledgments
We thank Wim Cosyn, Qintao Song, and Weihua Yang for valuable discussions.
This work was supported by the National Key R\&D Program of China No.~2024YFA1611004, by the National Natural Science Foundation of China (Grants Nos. 12175117, 12475084, and 12321005), by the Shandong Province Natural Science Foundation (Grants Nos. ZFJH202303 and ZR2024MA012), and by the European Union ``Next
Generation EU'' program through the Italian PRIN 2022 grant
No. 20225ZHA7W.

\appendix
\section{Hadronic tensor in terms of partonic functions}
\label{a.hadrontensor}
In the appendix, we present the complete expression of the tensor-polarized part of the hadronic tensor up to twist-3, 
\begin{align*}
    &W^{\mu\nu}=2z\sum_{a}e_{a}^{2}\int d^{2}\boldsymbol{p}_{T}d^{2}\boldsymbol{k}_{T}\delta^{2}(\boldsymbol{p}_{T}+\boldsymbol{q}_{T}-\boldsymbol{k}_{T})\\
    &\times\Biggl\{-g_T^{\mu\nu}S_{LL}f_{1LL}D_1 
        - \frac{p_{T}^{\{\mu}k_{T}^{\nu\}}-\left(k_{T}\cdot p_{T}\right)g_{T}^{\mu\nu}}{MM_{h}} S_{LL} h_{1LL}^\perp H_1^\perp \nonumber\\
        &+g_T^{\mu\nu} \frac{S_{LT}\cdot p_T}{M} f_{1LT}D_1
        -\frac{S_{LT}^{\{\mu}k_{T}^{\nu\}}-\left(S_{LT}\cdot k_{T}\right)g_{T}^{\mu\nu}}{M_h}h_{1LT}H_1^\perp\nonumber\\
        &+\bigg(\frac{p_{T}^{\{\mu}k_{T}^{\nu\}}-\left(k_{T}\cdot p_{T}\right)g_{T}^{\mu\nu}}{MM_h}\frac{S_{LT}\cdot p_T}{M}
        -\frac{S_{LT}^{\{\mu}k_{T}^{\nu\}}-\left(S_{LT}\cdot k_{T}\right)g_{T}^{\mu\nu}}{M_h}\frac{p_T^2}{2M^2} \bigg) h_{1LT}^\perp H_1^\perp\nonumber\\
        &-g_T^{\mu\nu}\frac{p_T\cdot S_{TT}\cdot p_T}{M^2} f_{1TT}D_1
        +\frac{k_T^{\{\mu}S_{TT}^{\nu\}\alpha} p_{T\alpha}-S_{TT}^{\alpha\beta}p_{T\alpha} k_{T\beta}g_T^{\mu\nu}}{M M_h}h_{1TT}H_1^\perp\nonumber\\
        &-\bigg(\frac{p_T^{\{\mu}k_T^{\nu\}}-(k_T\cdot p_T) g_T^{\mu\nu}}{MM_h}\frac{S_{TT}^{\alpha\beta}p_{T\alpha} p_{T\beta}}{M^2}
        -\frac{k_T^{\{\mu}S_{TT}^{\nu\}\alpha}p_{T\alpha}-S_{TT}^{\alpha\beta}p_{T\alpha} k_{T\beta}g_T^{\mu\nu}}{MM_h}\frac{p_T^2}{2M^2}\bigg)h_{1TT}^\perp H_1^\perp\nonumber\\
        &+ i\frac{S_{LT}^{[\mu}p_T^{\nu]}}{M}g_{1LT}D_1 + i\frac{S_{TT}^{\alpha[\mu}p_T^{\nu]}p_{T\alpha}}{M^2}g_{1TT}D_1\\
    &+\frac{2M}{Q}\frac{\hat{t}^{\{\mu}k_{T}^{\nu\}}}{M_{h}}S_{LL}xh_{LL}H_{1}^{\perp}+\frac{2}{Q}\hat{t}^{\{\mu}p_{T}^{\nu\}}S_{LL}xf_{LL}^{\perp}D_{1}+\frac{2M}{Q}\hat{t}^{\{\mu}S_{LT}^{\nu\}}xf_{LT}D_{1}\\
    &+\bigg(\frac{2}{Q}\hat{t}^{\{\mu}S_{LT}^{\nu\}}\frac{k_{T}\cdot p_{T}}{M_{h}}-\frac{2}{Q}\hat{t}^{\{\mu}p_{T}^{\nu\}}\frac{S_{LT}\cdot k_{T}}{M_{h}}\bigg)xh_{LT}^{\perp}H_{1}^{\perp}+\frac{2}{Q}\frac{\hat{t}^{\{\mu}k_{T}^{\nu\}}}{M_{h}}(S_{LT}\cdot p_{T})xh_{LT}H_{1}^{\perp}\\
    &+\bigg(\frac{1}{Q}\hat{t}^{\{\mu}S_{LT}^{\nu\}}\frac{p_{T}^{2}}{M}-\frac{2}{Q}\hat{t}^{\{\mu}p_{T}^{\nu\}}\frac{S_{LT}\cdot p_{T}}{M}\bigg)xf_{LT}^{\perp}D_{1}\\
    &-\frac{2}{Q}\hat{t}^{\{\mu}S_{TT}^{\nu\}\rho}p_{T\rho}xf_{TT}D_{1}+\bigg(\frac{2}{Q}\hat{t}^{\{\mu}p_{T}^{\nu\}}\frac{k_{T}\cdot S_{TT}\cdot p_{T}}{MM_{h}}-\frac{2}{Q}\hat{t}^{\{\mu}S_{TT}^{\nu\}\rho}p_{T\rho}\frac{k_{T}\cdot p_{T}}{MM_{h}}\bigg)xh_{TT}^{\perp}H_{1}^{\perp}\\
    &-\frac{2}{Q}\frac{\hat{t}^{\{\mu}k_{T}^{\nu\}}}{M_{h}}\frac{p_{T}\cdot S_{TT}\cdot p_{T}}{M^2}xh_{TT}H_{1}^{\perp}+\bigg(\frac{2}{Q}\hat{t}^{\{\mu}p_{T}^{\nu\}}\frac{p_{T}\cdot S_{TT}\cdot p_{T}}{M^{2}}-\frac{1}{Q}\hat{t}^{\{\mu}S_{TT}^{\nu\}\rho}p_{T\rho}\frac{p_{T}^{2}}{M^{2}}\bigg)xf_{TT}^{\perp}D_{1}\\
    &-i\frac{2M}{Q}\frac{\hat{t}^{[\mu}k_{T}^{\nu]}}{M_{h}}S_{LL}xe_{LL}H_{1}^{\perp}+i\frac{2M}{Q}\frac{\hat{t}^{[\mu}p_{T}^{\nu]}}{M}S_{LL}xg_{LL}^{\perp}D_{1}+i\frac{2M}{Q}\hat{t}^{[\mu}S_{LT}^{\nu]}xg_{LT}D_{1}\\
    &+\bigg(i\frac{2}{Q}\hat{t}^{[\mu}p_{T}^{\nu]}\frac{S_{LT}\cdot k_{T}}{M_{h}}-i\frac{2}{Q}\hat{t}^{[\mu}S_{LT}^{\nu]}\frac{k_{T}\cdot p_{T}}{M_{h}}\bigg)xe_{LT}^\perp H_{1}^{\perp}+i\frac{2}{Q}\frac{\hat{t}^{[\mu}k_{T}^{\nu]}}{M_{h}}(S_{LT}\cdot p_{T})xe_{LT} H_{1}^{\perp}\\
    &+\bigg(i\frac{1}{Q}\hat{t}^{[\mu}S_{LT}^{\nu]}\frac{p_{T}^{2}}{M}-i\frac{2}{Q}\hat{t}^{[\mu}p_{T}^{\nu]}\frac{S_{LT}\cdot p_{T}}{M}\bigg)xg_{LT}^{\perp}D_{1}-i\frac{2}{Q}\hat{t}^{[\mu}S_{TT}^{\nu]\rho}p_{T\rho}xg_{TT}D_{1}\\
    &+\bigg(i\frac{2}{Q}\hat{t}^{[\mu}S_{TT}^{\nu]\rho}p_{T\rho}\frac{k_{T}\cdot p_{T}}{MM_{h}}-i\frac{2}{Q}\hat{t}^{[\mu}p_{T}^{\nu]}\frac{k_{T}\cdot S_{TT}\cdot p_{T}}{MM_{h}}\bigg)xe_{TT}^\perp H_{1}^{\perp}\\
    &-i\frac{2}{Q}\frac{\hat{t}^{[\mu}k_{T}^{\nu]}}{M_{h}}\frac{p_{T}\cdot S_{TT}\cdot p_{T}}{M}xe_{TT} H_{1}^{\perp}\\
    &+\bigg(i\frac{2}{Q}\hat{t}^{[\mu}p_{T}^{\nu]}\frac{p_{T}\cdot S_{TT}\cdot p_{T}}{M^{2}}-i\frac{1}{Q}\hat{t}^{[\mu}S_{TT}^{\nu]\rho}p_{T\rho}\frac{p_{T}^{2}}{M^{2}}\bigg)xg_{TT}^{\perp}D_{1}\\
    &+\frac{2}{Q}\hat{t}^{\{\mu}k_{T}^{\nu\}}S_{LL}f_{1LL}\frac{\tilde{D}^{\perp}}{z}+\frac{2}{Q}\hat{t}^{\{\mu}p_{T}^{\nu\}}S_{LL}\frac{M_{h}}{M}h_{1LL}^{\perp}\frac{\tilde{H}}{z}\\&+\frac{2}{Q}\hat{t}^{\{\mu}S_{LT}^{\nu\}}M_{h}h_{1LT}\frac{\tilde{H}}{z}+\bigg(\frac{2}{Q}\hat{t}^{\{\mu}S_{LT}^{\nu\}}\frac{k_{T}\cdot p_{T}}{M}-\frac{2}{Q}\hat{t}^{\{\mu}p_{T}^{\nu\}}\frac{M_{h}}{M}\frac{S_{LT}\cdot k_{T}}{M_{h}}\bigg)g_{1LT}\frac{\tilde{G}^{\perp}}{z}\\
    &-\frac{2}{Q}\hat{t}^{\{\mu}k_{T}^{\nu\}}\frac{S_{LT}\cdot p_{T}}{M}f_{1LT}\frac{\tilde{D}^{\perp}}{z}+\bigg(\frac{1}{Q}\hat{t}^{\{\mu}S_{LT}^{\nu\}}M_{h}\frac{p_{T}^{2}}{M^{2}}-\frac{2}{Q}\frac{\hat{t}^{\{\mu}p_{T}^{\nu\}}}{M}M_{h}\frac{S_{LT}\cdot p_{T}}{M}\bigg)h_{1LT}^{\perp}\frac{\tilde{H}}{z}\\&-\frac{2}{Q}\frac{\hat{t}^{\{\mu}S_{TT}^{\nu\}\rho}p_{T\rho}}{M}M_{h}h_{1TT}\frac{\tilde{H}}{z}+\bigg(\frac{2}{Q}\frac{\hat{t}^{\{\mu}S_{TT}^{\nu\}\rho}p_{T\rho}}{M}\frac{k_{T}\cdot p_{T}}{M}-\frac{2}{Q}\frac{\hat{t}^{\{\mu}p_{T}^{\nu\}}}{M}\frac{k_{T}\cdot S_{TT}\cdot p_{T}}{M}\bigg)g_{1TT}\frac{\tilde{G}^{\perp}}{z}\\&+\frac{2}{Q}\hat{t}^{\{\mu}k_{T}^{\nu\}}\frac{p_{T}\cdot S_{TT}\cdot p_{T}}{M^{2}}f_{1TT}\frac{\tilde{D}^{\perp}}{z}+\bigg(-\frac{1}{Q}\frac{\hat{t}^{\{\mu}S_{TT}^{\nu\}\rho}p_{T\rho}}{M}\frac{p_{T}^{2}}{M^{2}}M_{h}+\frac{2}{Q}\hat{t}^{\{\mu}p_{T}^{\nu\}}\frac{p_{T}\cdot S_{TT}\cdot p_{T}}{M^{3}}M_{h}\bigg)h_{1TT}^{\perp}\frac{\tilde{H}}{z}\\
    &-i\frac{2}{Q}\hat{t}^{[\mu}k_{T}^{\nu]}S_{LL}f_{1LL}\frac{\tilde{G}^{\perp}}{z}+i\frac{2}{Q}\frac{\hat{t}^{[\mu}p_{T}^{\nu]}}{M}M_{h}S_{LL}h_{1LL}^{\perp}\frac{\tilde{E}}{z}\\
    &+i\frac{2}{Q}\hat{t}^{[\mu}S_{LT}^{\nu]}M_{h}h_{1LT}\frac{\tilde{E}}{z}+\bigg(i\frac{2}{Q}\hat{t}^{[\mu}S_{LT}^{\nu]}\frac{k_{T}\cdot p_{T}}{M}-i\frac{2}{Q}\frac{\hat{t}^{[\mu}p_{T}^{\nu]}}{M}S_{LT}\cdot k_{T}\bigg)g_{1LT}\frac{\tilde{D}^{\perp}}{z}\\
    &+i\frac{2}{Q}\hat{t}^{[\mu}k_{T}^{\nu]}\frac{S_{LT}\cdot p_{T}}{M}f_{1LT}\frac{\tilde{G}^{\perp}}{z}+\bigg(i\frac{1}{Q}\hat{t}^{[\mu}S_{LT}^{\nu]}\frac{p_{T}^{2}}{M^{2}}M_{h}-i\frac{2}{Q}\frac{\hat{t}^{[\mu}p_{T}^{\nu]}}{M}\frac{S_{LT}\cdot p_{T}}{M}M_{h}\bigg)h_{1LT}^{\perp}\frac{\tilde{E}}{z}\\&-i\frac{2}{Q}\frac{\hat{t}^{[\mu}S_{TT}^{\nu]\rho}p_{T\rho}}{M}M_{h}h_{1TT}\frac{\tilde{E}}{z}+\bigg(i\frac{2}{Q}\frac{\hat{t}^{[\mu}S_{TT}^{\nu]\rho}p_{T\rho}}{M}\frac{k_{T}\cdot p_{T}}{M}-i\frac{2}{Q}\frac{\hat{t}^{[\mu}p_{T}^{\nu]}}{M}\frac{k_{T}\cdot S_{TT}\cdot p_{T}}{M}\bigg)g_{1TT}\frac{\tilde{D}^{\perp}}{z}\\&-i\frac{2}{Q}\hat{t}^{[\mu}k_{T}^{\nu]}\frac{p_{T}\cdot S_{TT}\cdot p_{T}}{M^{2}}f_{1TT}\frac{\tilde{G}^{\perp}}{z}+\bigg(-i\frac{1}{Q}\frac{\hat{t}^{[\mu}S_{TT}^{\nu]\rho}p_{T\rho}}{M}\frac{p_{T}^{2}}{M^{2}}M_{h}+i\frac{2}{Q}\frac{\hat{t}^{[\mu}p_{T}^{\nu]}}{M}\frac{p_{T}\cdot S_{TT}\cdot p_{T}}{M^{2}}M_h\bigg)h_{1TT}^{\perp}\frac{\tilde{E}}{z}\Bigg\}.
    \end{align*}

\end{document}